\documentclass[epj]{svjour}
\usepackage{graphicx,epsfig,rotating,multirow,amssymb,url}
\usepackage{subfig,psfrag}

\begin{document}

\newcommand{\ra}{\rightarrow}
\newcommand{\lsim}{\mbox{\raisebox{-.6ex}{~$\stackrel{<}{\sim}$~}}}
\newcommand{\gsim}{\mbox{\raisebox{-.6ex}{~$\stackrel{>}{\sim}$~}}}



\def\makeheadbox{\flushright{{HIP-2008-25/TH}\\{DO-TH 08/04}}}

\title{Search for Higgs Bosons in SUSY Cascades in CMS and 
  Dark Matter with Non-universal Gaugino Masses}

\titlerunning{Higgs Bosons and Dark Matter with Non-universal Gaugino
  Masses}

\author{Katri Huitu \inst{1} \fnmsep \inst{2} \fnmsep \thanks{katri.huitu@helsinki.fi}
  \and Ritva Kinnunen\inst{1} \fnmsep \thanks{ritva.kinnunen@helsinki.fi}
  \and Jari Laamanen \inst{3} \fnmsep \inst{1}
  \fnmsep \thanks{jari.laamanen@uni-dortmund.de}%
  \and Sami Lehti \inst{1} \fnmsep \thanks{sami.lehti@helsinki.fi}
  \and Sourov Roy \inst{4} \fnmsep \thanks{tpsr@iacs.res.in}
  \and Tapio Salminen \inst{2} \fnmsep \thanks{Tapio.Salminen@helsinki.fi}%
}
\institute{Helsinki Institute of Physics, P.O.~Box 64, 
  {FIN-00014} University of Helsinki, Finland
  \and
  Department of Physics, P.O.~Box 64, 
  {FIN-00014} University of Helsinki, Finland
  \and
  Institut f{\"u}r Physik,
  Technische Universit{\"a}t Dortmund,
  {D-44221} Dortmund, Germany
  \and
  Department of Theoretical Physics
  and Centre for Theoretical Sciences,\\
  Indian Association for the Cultivation of Science,
  2A \& 2B Raja S.C. Mullick Road,
  Kolkata {700 032}, India
}
%
\date{}
\abstract{In grand unified theories (GUT), non-universal boundary conditions for the 
gaugino masses may arise at the unification scale, and
affect the observability of the neutral
MSSM Higgs bosons (h/H/A) at the LHC. 
The implications of such non-universal gaugino masses
are investigated for the Higgs boson production in the SUSY cascade
decay chain $\tilde{g} \rightarrow \tilde{q}q$, $\tilde{q} \rightarrow  \tilde{\chi_2}q$,
$ \tilde{\chi_2} \rightarrow \tilde{\chi_1} h/H/A$, $h/H/A  \rightarrow b\bar{b}$ produced in
pp interactions. In the singlet representation
with universal gaugino masses only the light Higgs boson can be
produced in this cascade with the parameter region of interest for us,
while with non-universal gaugino masses 
heavy neutral MSSM Higgs boson production may dominate.
The allowed parameter space in the light of the WMAP constraints on the 
cold dark matter relic density is investigated
in the above scenarios for gaugino mass parameters.
We also demonstrate that combination of representations can give the
required amount of dark matter in any point of the parameter space.
In the non-universal case we show
that heavy Higgs bosons can be detected in the studied cascade
in parameter regions with the WMAP 
preferred neutralino relic density.
\PACS{{12.60.Jv}{Supersymmetric models} 
  \and {95.35.+d}{Dark matter}
  \and {14.80.Cp}{Non-standard-model Higgs bosons}
  \and {14.80.Ly}{Supersymmetric partners of known particles}
  \and {12.10.Kt}{Unification of couplings; mass relations}
}
} 

\maketitle

\section{Introduction}

Most studies concerning the discovery potential for finding
supersymmetric Higgs bosons at the LHC, have been done in models with
universal gaugino masses at the unification scale. There is, however,
no compelling theoretical reason for such a choice. In grand unified
supersymmetric models, which include an ${\rm SU(5)}$ grand unified
model, non-universal gaugino masses are generated by a nonsinglet
chiral superfield $\Phi^n$ that appears linearly in the gauge kinetic
function $f(\Phi)$. The function $f(\Phi)$ is an analytic function of
the chiral superfields $\Phi$ in the theory \cite{Cremmer:1982wb}. It
should be noted that the chiral superfields $\Phi$ consist of a set of
gauge singlet superfields $\Phi^s$ and gauge nonsinglet superfields
$\Phi^n$, respectively, under the grand unified group. If the
auxiliary part $F_\Phi$ of a chiral superfield $\Phi$ in the $f(\Phi)$
gets a VEV, then gaugino masses arise from the coupling of $f(\Phi)$
with the field strength superfield $W^a$. The Lagrangian for the
coupling of gauge kinetic function with the gauge field strength is
written as
\begin{eqnarray}
{\cal L}_{gk} = \int d^2 \theta f_{ab}(\Phi) W^a W^b + H.c.
\label{gauge-kinetic}
\end{eqnarray} 
where $a$ and $b$ are gauge group indices [for example, $a,b =
1,2,...,24$ for ${\rm SU(5)}$], and repeated indices are summed
over. The gauge kinetic function $f_{ab}(\Phi)$ is
\begin{eqnarray}
f_{ab}(\Phi) = f_0(\Phi^s)\delta_{ab} + \sum_n f_n(\Phi^s) {\Phi^n_{ab} \over
M_P} + \cdots, 
\end{eqnarray}
where as described above the $\Phi^s$ and $\Phi^n$ are the singlet and
nonsinglet chiral superfields, respectively. Here $f_0(\Phi^s)$ and
$f_n(\Phi^s)$ are functions of gauge singlet superfields $\Phi^s$, and
$M_P$ is some large scale. When $F_\Phi$ gets a VEV $\langle F_\Phi
\rangle$, the interaction (\ref{gauge-kinetic}) gives rise to gaugino
masses$\colon$
\begin{eqnarray}
{\cal L}_{gk} \supset {{\langle F_\Phi \rangle}_{ab} \over M_P}\lambda^a
\lambda^b + H.c., 
\end{eqnarray}
where $\lambda^{a,b}$ are gaugino fields. It should be noted that the
${\rm U(1)}$, ${\rm SU(2)}$, and ${\rm SU(3)}$ gauginos are denoted by
$\lambda_1$, $\lambda_2$, and $\lambda_3$, respectively.

Since the gauginos belong to the adjoint representation of the gauge
group, in the case of ${\rm SU(5)}$ for example, $\Phi$ and $F_\Phi$
can belong to any of the following representations appearing in the
symmetric product of the two {\bf 24} dimensional representations of
${\rm SU(5)}\colon$
\begin{eqnarray}
{({\mathbf {24}} \otimes {\mathbf {24}})}_{Symm} = {\mathbf 1} \oplus
{\mathbf {24}} \oplus {\mathbf {75}} \oplus {\mathbf {200}}.
\end{eqnarray}
In the minimal case (which is the simplest one too), $\Phi$ and
$F_\Phi$ are assumed to be in the singlet representation of ${\rm
  SU(5)}$. This corresponds to equal gaugino masses at the GUT
scale. However, $\Phi$ can belong to any of the nonsinglet
representations ${\mathbf {24}}$, ${\mathbf {75}}$, and ${\mathbf
  {200}}$ of ${\rm SU(5)}$. In that case, the gaugino masses are
unequal but related to one another via the representation invariants
\cite{Ellis:1985jn,Drees:1985bx,Anderson:1996bg}.  It should be kept
in mind that an arbitrary combination of these different
representations is also allowed.  We shall first study the case of
each representation separately, and then the effect of combining two
of these.  In Table \ref{tab:gaug} we display the ratios of resulting
gaugino masses at tree level as they arise when $F_\Phi$ belongs to
various representations of ${\rm SU(5)}$. Clearly, the nonsinglet
representations have characteristic mass relationships for the
gauginos at the GUT scale. The resulting relations at the electroweak
scale, using the renormalization group (RG) evolution at the one-loop
level are also displayed.
\begin{table}[h]
\small
\centering
\vskip 0.1 in
\caption{Ratios of gaugino mass parameters at the GUT scale in the
  normalization $M_3 ({\rm GUT})$ = 1 and at the electroweak scale in
  the normalization $M_3 ({\rm EW})$ = 1 at the one-loop level.}
\begin{tabular}{|l|ccc|rrc|}
  \hline \hline
$F_\Phi$  &  $M^G_1$ & $M^G_2$ & $M^G_3$ & $M^{\rm EW}_1$ & $M^{\rm EW}_2$ 
& $M^{\rm EW}_3$ \\
  \hline
$\mathbf{1}$   & 1 & 1 & 1 &0.14 & 0.29 & 1\\
$\mathbf{24}$  &--0.5 &--1.5 &1 &--0.07 & --0.43 & 1\\
$\mathbf{75}$  &--5  &3  &1  &--0.72 & 0.87 & 1\\
$\mathbf{200}$ &10  &2  &1  &1.44 & 0.58 & 1\\
  \hline
  \hline
\end{tabular}
\label{tab:gaug}
\end{table}

The phenomenology of supersymmetric models with nonuniversal gaugino
masses has been considered e.g. in
\cite{Huitu:2005wh,Anderson:2000,Huitu:2000,Belanger:2005,Djouadi:2001,Chattopadhyay:2005,Chattopadhyay:2001mj,Bhattacharya:2007dr}.
Studies were performed also in the context of supersymmetric dark
matter
\cite{Corsetti:2001,Bertin:2003,Birkedal-Hansen:2003,Chattopadhyay:2003,Chattopadhyay:2001va,Cerdeno:2004zj,King:2007vh}.
The phenomenology of supersymmetric models depends crucially on the
compositions of neutralinos and charginos.  Hence, it is extremely
important to investigate the changes in the experimental signatures
with the changes in the composition of neutralinos and charginos which
arise because of the non-universal gaugino masses at the GUT scale. In
this study our focus will be on the experimental signatures of neutral
Higgs bosons appearing in the cascade decay chains of the squark and
gluinos produced at the LHC.  If the squarks and gluinos are light
enough, their production cross sections are large at the LHC. The
light neutralinos $\tilde{\chi}_{1,2}$ are typical decay products of
$\tilde{\rm g}$ and $\tilde{\rm q}$. The neutral Higgs bosons can be
produced in the decay of $\tilde{\chi}_{2}$, if the mass difference
between $\tilde{\chi}_{2}$ and $\tilde{\chi}_{1}$ is large enough.  As
the production rate is largely independent of the value of tan$\beta$,
these production channels have been found particularly interesting at
the LHC to cover the difficult region of low and medium tan$\beta$
values \cite{Baer:1992,Datta:2003iz}.  Production of Higgs via
$\tilde{\chi_2}\ra \tilde{\chi_1} \rm h/H/A$ in models where gauginos
are in singlet and {\bf 24} representation were studied in
\cite{Huitu:2005wh}.  Also, recently the role of gaugino mass
non-universality (in a model independent framework) in the context of
Higgs productions has been discussed in \cite{arXiv:0806.2367}.  It
was found for sample parameters that only the lightest Higgs could be
produced in the model with singlet, while also the heavy Higgses, H
and A, could be produced from the cascade in the model with {\bf 24}
representation.  The gaugino mass parameters indeed indicate that it
is more probable to have the cascade decay including heavy Higgses in
the model with gauginos in {\bf 24}.  Namely, if the $\mu$ parameter,
mixing Higgs superfields in the MSSM superpotential, is large, the
lightest neutralino using either representations is close to bino,
while the second lightest is close to wino.  From Table 1, in the
model with singlet the mass difference of the neutralinos is then
close to the mass of the lightest neutralino, since $m_{\chi_2}\simeq
2m_{\chi_1}$, while in the model with {\bf 24}, $m_{\chi_2}\simeq
6m_{\chi_1}$ and the mass difference thus five times the mass of the
lightest neutralino.  Note that the cascade is not possible with small
values of the $\mu$-parameter, since in such a case $\tilde\chi_2$ and
$\tilde\chi_1$ are similar in mass.  From Table 1, one can also note
that the cascade in the cases of representation {\bf 75} is not
probable.

In this work, the Higgs boson decay to b quarks, \linebreak $\rm H/A/h
\to b\bar b$, with a large branching fraction ($\sim$0.9) is
selected. In the multi-jet events this decay mode suffers from a large
combinatorial background, but structuring the event into two
hemispheres according to the directions of the decaying gluinos and
squarks from the hard process, this background can be efficiently
reduced.  Production of the light Higgs boson h in SUSY cascades with
$\rm b\bar{\rm b}$ final state has been studied in
\cite{CMSNote2006/090} in the CMS detector with the assumption of
universal gaugino masses. The event selection is based on a
requirement of four energetic jets, large missing $\rm E_{\rm T}$,
separation of the jets into two hemispheres and the reconstruction of
the Higgs boson mass from two jets tagged as b jets.  Similar analysis
method is used here. The main phenomenological differences between
approaches of Ref.~\cite{CMSNote2006/090} and this work are the
different branching fractions of the SUSY particles. In the singlet
representation only the light Higgs boson $\rm h$ can be produced
whereas in the representation {\bf 24} the $\tilde{\chi}_{2} \ra \rm
H/A \tilde{\chi}_{1}$ decays dominate in most of the parameter
space. Therefore considerable differences can be expected in the final
mass distribution as well as the obtainable significance of the
signal. In Ref.~\cite{Huitu:2005wh} the 4b final state from H/A/h~$\ra
\rm b\bar{\rm b}$ decay from both of the cascades has been proposed as
the search channel. This mode however is subject to a large
combinatorial background from $\rm b\bar{\rm b}$ combinations due to
real or fake b jets in the cascade not containing the Higgs boson.

Combining information from different sources is essential for
utilizing future facilities in best possible way.  In addition to the
laboratory studies, relevant input is obtained from the dark matter
searches: the WMAP satellite has put precise limits on the relic
density.  Supersymmetric theories which preserve $R$ parity contain a
natural candidate for the cold dark matter particle.  If the lightest
neutralino is the lightest supersymmetric particle (LSP), it can
provide the appropriate relic density.
In many SUGRA type models the lightest neutralino is bino-like, which
often leads to too large thermal relic density, as compared to the
limits provided by the WMAP experiment
\cite{Spergel:2003cb,Spergel:2006hy}.
Of the possible representations of SU(5) GUT all but the singlet
representation produce nonuniversal gaugino masses.  Thus the
resulting neutralino composition changes from the usual universal
gaugino mass case \cite{Huitu:2005wh}.  Depending on the
representation where the gauginos are, this leads to different regions
in the parameter space where WMAP constraints are satisfied.  It is
also shown that by combining representations, observationally allowed
dark matter parameter regions are not very restricted, if parameters
are suitably chosen.

Here we are interested in the case when both the relic density
constraints from WMAP can be satisfied and the heavy Higgses can be
produced in the cascade.  An important contribution to the neutralino
annihilation comes from the enhancement due to the Higgs resonance.
If we denote the mass of the heavy Higgs, either H or A, by $m_A$, the
enhanced annihilation occurs when $m_{\tilde\chi_1}=0.5 m_A$.  Since
the mass difference between the two lightest neutralinos is $0.5 m_A$
in the model with singlet and $2.5 m_A$ in the model with {\bf 24}, it
is obvious that this mechanism is available in the latter case, but
not in the case with singlet, if we want to study the Higgs production
via cascade at the same time.  Indeed, we find easily parameter sets
with both neutralino dark matter and the cascade to all the Higgses h,
H, and A in the model with {\bf 24 } but not in the model with
singlet.  Thus in this work we are especially interested in the model
with representation {\bf 24}, for which also the production cross
sections for heavy Higgses were found reasonably large in
\cite{Huitu:2005wh}.  In order to have both relevant region for dark
matter and to produce heavy Higgses in the cascade decays, the model
with {\bf 24} is special among those with nonuniversal SU(5)
representations.  We shall see later that in the model with {\bf 200},
the relic density is never large enough to produce the dark matter
solely.  In {\bf 75} the mass difference between the two lightest
neutralinos is too small for the cascade decay in the parameter region
with relic density in the WMAP region.

In the following we first discuss in section 2 the parameter regions
for observationally acceptable dark matter in different SU(5)
representations.  In section 3 we concentrate on representation {\bf
  24} and specifically study the parameter region, where the Higgs
production from neutralino decay is most promising, and the signal and
background processes are discussed.  The detector simulation is
shortly presented in Section 4.  The event selection and the signal to
background ratio are discusssed in Section 5, and the background
determination is elaborated in Section 6.  Section 7 describes the
systematic uncertainties on the background determination. Discovery
potential is given in Section 8 and the conclusions in Section 9.

\section{Dark matter in SU(5) representations}

We start the description of the relic density for four different SU(5)
representations with the representation \textbf{24}, which is most
interesting for us here, as explained in the introduction. We then
shortly discuss the representations \textbf{75} and \textbf{200}, and
compare with the singlet, which has been studied extensively in the
literature.  Then we discuss combination of singlet representation
with \textbf{200}.

\begin{figure*}
\psfrag{m2}{$M_{2}$}
\psfrag{m0}{$m_{0}$}
 \centering 
 \subfloat[$\tan\beta = 10,\ \rm{sgn}(\mu) = +1,\ A_0=0$]{
    \label{fig:rd24-a}
   \includegraphics[width=0.46\textwidth]{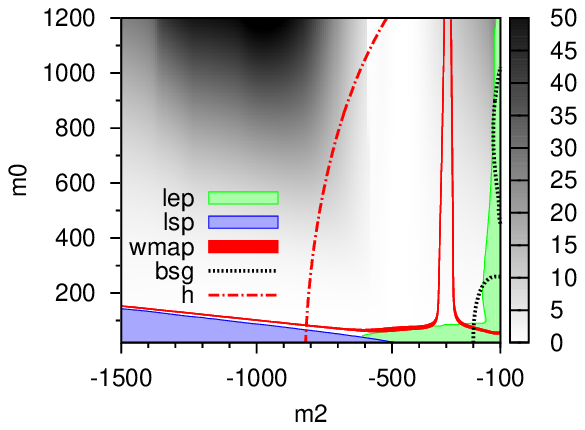}}
 \hspace{3mm} 
 \subfloat[$\tan\beta = 35,\ \rm{sgn}(\mu) = +1,\ A_0=0$]{
   \label{fig:rd24-b}
   \includegraphics[width=0.46\textwidth]{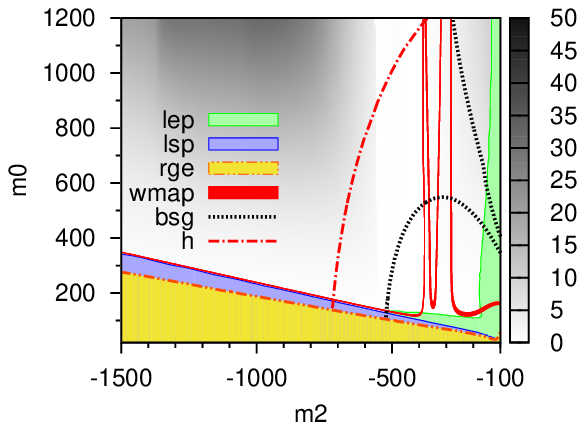}}\\
 \subfloat[$\tan\beta = 10,\ \rm{sgn}(\mu) = -1,\ A_0 = 0$]{
   \label{fig:rd24-c}
   \includegraphics[width=0.46\textwidth]{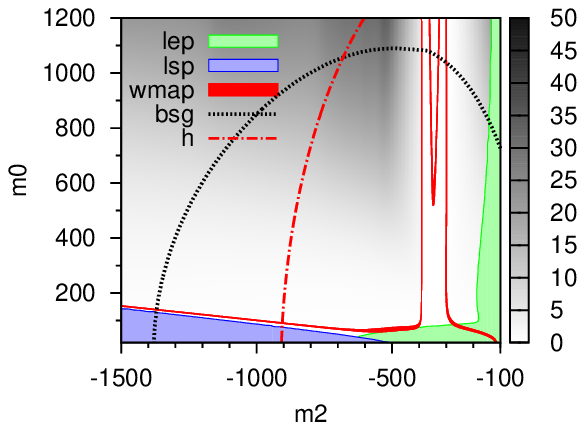}}
 \hspace{3mm} 
 \subfloat[$\tan\beta = 10,\ \rm{sgn}(\mu) = -1,\ A_0 = 1$ TeV]{
   \label{fig:rd24-d}
   \includegraphics[width=0.46\textwidth]{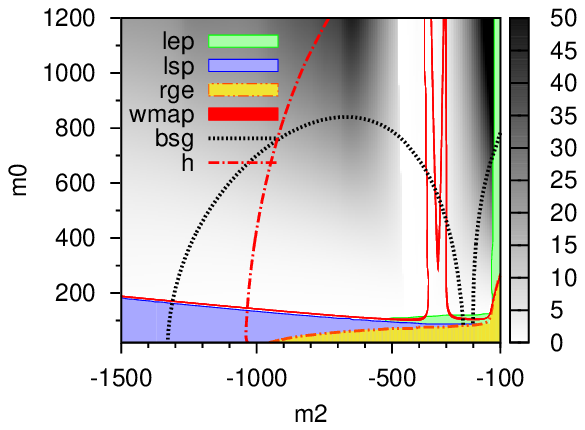}}
 \caption{Relic density $\Omega_\chi h^2$ in the representation
   \textbf{24} in ($M_{2}, m_0$) plane. The
   dark shaded areas represent the larger relic density.  \textsf{wmap}
   denoted filling
   is the WMAP preferred region,
   \textsf{lep} shows an area next to the axes, where the experimental
   mass limits are not met, \textsf{rge} shows an area where there is
   no radiative EWSB, and \textsf{lsp} the area where neutralino is not 
   the LSP.
   \textsf{h} gives the line of $m_h = 114$ GeV.
    The area between two \textsf{bsg} contours is allowed by
   $b \to s \gamma$ limits in (a), (b),
   and (d), and above the contour in (c).}
 \label{fig:relic24}
\end{figure*}

The amount of thermal relic density in the representation \textbf{24}
is presented in Fig.~\ref{fig:relic24} for four different sets of GUT
scale parameters.
The spectrum was calculated with {\tt SOFTSUSY} \cite{Allanach:2001kg}
and relic densities and constraints with {\tt micrOMEGAs}
\cite{Belanger:2001fz,Belanger:2004yn,Belanger:2006is}.
The dark shaded areas represent larger relic density than the lighter
areas.  For the relic density, we use here the WMAP combined three
year limits \cite{Spergel:2003cb,Spergel:2006hy}
\begin{eqnarray}
  \Omega_{CDM} h^2 = 0.11054^{+0.00976}_{-0.00956} \quad (2\sigma).  
\end{eqnarray}
For the particle masses, the following limits are applied \cite{Belanger:2006is}:
$m_{\tilde e_R} > 99.4$ or 100.5 GeV depending if the lightest
neutralino mass is below or above 40 GeV, 
$m_{\tilde \mu_R} > 95$ GeV, 
$m_{\tilde\tau_1} > 80.5$ to 88 GeV depending on the lightest
neutralino mass (from 10 to 75 GeV), 
$m_{\tilde \nu_i} > 43$ GeV, and 
$m_{\tilde\chi^\pm} > 73.1$ to 103 GeV depending on the sneutrino
masses (from 45 to 425 GeV).
The curve $m_h=114$ GeV is depicted in the figure.  For the shown
parameter region, when otherwise experimentally allowed, Higgs is
always heavier than $91$ GeV, which is the Higgs mass limit in MSSM
for $\tan\beta \geq 10$ assuming maximal top mixing \cite{unknown:2001xy}.

The most striking feature in Fig.~\ref{fig:relic24} is the valley of
the low relic density area around $M_2 \sim -300$ GeV. There the
annihilation of neutralinos proceeds to quarks and the minimum occurs
at the $Z$ peak providing an efficient annihilation.
Outside of the valley the relic density rises, overclosing the universe.
In the representation \textbf{24} the lightest neutralino is very
bino-like, and the WMAP preferred region tends to be quite narrow.

If the overall relic density is low, the crucial differences that
control the preferred regions of the parameter space come mostly from
the composition of the lightest neutralino.  An important single
aspect that separates the different parameter sets in
Fig.~\ref{fig:relic24} is the presence of the higgsino component in
the lightest neutralino, which increases annihilation into the gauge
boson pairs.
Increasing the value of $\tan\beta$ decreases the value of the
supersymmetric higgsino mass parameter $\mu$, as can be seen from the
electroweak symmetry breaking (EWSB) condition
\begin{equation}
  |\mu|^2 = \frac{m^2_{H_d} - m^2_{H_u} \tan^2\beta}{\tan^2\beta -1}
  -\frac 1 2 m^2_{Z}, \label{mzminim}
\end{equation}
where $\tan\beta \equiv \langle H^0_u \rangle / \langle H^0_d \rangle
= v_u/v_d$.
This results in a bigger higgsino component in the lightest
neutralino, which then annihilates more efficiently.  The effect is
seen in comparing Figures \ref{fig:rd24-a} and \ref{fig:rd24-b}.
Since the relic density in Fig.~\ref{fig:rd24-b} is smaller than in
Fig.~\ref{fig:rd24-a} in general, also annihilation to the lightest
Higgs suffices to lower the relic density to the WMAP preferred
region.  Thus the low relic density region is wider than in
Fig.~\ref{fig:rd24-a} with bands corresponding to both Z and h.  The
higher $\tan\beta$ also increases the area at the low $m_0$, where the
radiative EWSB is not possible (marked \textit{rge}, below the area
where neutralino is not the LSP).  \footnote{It can be noted that the
  Higgs funnel region, which is present in singlet case, does not
  exist in {\bf 24} representation at large $\tan\beta$ and large
  $m_0$ values.  This is due to the neutralinos, which are too light
  to annihilate to the heavy Higgs bosons at the resonance.}

The same mutual sign for the $\mu$ parameter and the gaugino mass
$M_1$ gives a slightly lower mass for the lightest neutralino than in
the case with opposite sign. Since in the representation \textbf{24}
the signs of $M_1$ and $M_2$ are the same, this effect can be seen
comparing figures \ref{fig:rd24-a} and \ref{fig:rd24-c}.  Thermal
relic density is proportional to the mass of the particle. The $b\to
s\gamma$ constraint is also prominent in the negative $\mu$ case
\cite{Nath:1994tn,Nath:1994ci,Baer:1997jq,Carena:2000uj,Degrassi:2000qf}.
We have used the two sigma world average of $BR(b\to s \gamma) = (355
\pm 24^{+9}_{-10} \pm 3) \times 10^{-6}$ for the branching fraction
\cite{Barberio:2007cr}.
For example in Fig.~\ref{fig:rd24-c} the visible area is mostly
excluded by the 2$\sigma$ limit. In other figures the $b\to s \gamma$
allowed region is between the \textit{bsg}-denoted contours.

Switching on the trilinear scalar coupling parameter $A_0$ increases
the value of $\mu$ through the RG evolution of the squared soft Higgs
mass parameters, therefore reducing the higgsino component in the
lightest neutralino. This can be seen comparing Figures
\ref{fig:rd24-c} and \ref{fig:rd24-d} at the low $|M_2|$ values. In
the Fig.~\ref{fig:rd24-c} the higgsino component of the LSP is larger
at the low $|M_2|$-high $m_0$ region, so the resulting relic density
is lower than in Fig.~\ref{fig:rd24-d}, where the lightest neutralino
is almost purely bino.\footnote{In Fig.~\ref{fig:rd24-d} the relic
  density can rise up to the values of 80; the shading reaches maximum
  saturation already in 50 in order to ease comparison with other
  figures.}

\begin{figure}
\psfrag{m2}{$M_{2}$}
\psfrag{m0}{$m_{0}$}
  \centering 
\includegraphics[width=0.5\textwidth,height=0.5\textwidth]{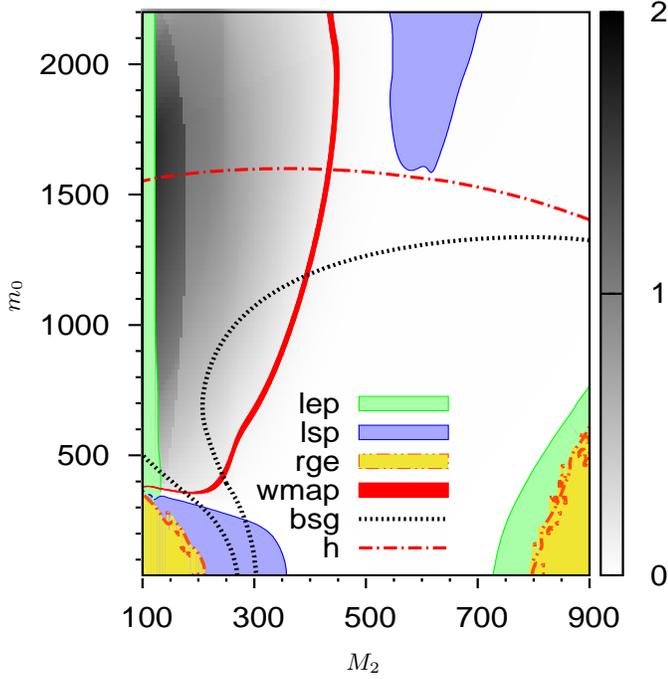}
  \caption{Relic density $\Omega_\chi h^2$ in the representation
    \textbf{75} in ($M_{2}, m_0$) plane.
    Explanation for the key is as in Fig.~\ref{fig:relic24}.
    The parameters used here are $\tan\beta = 10,\ \rm{sgn}(\mu) =
    -1,\ A_0 = 1$ TeV.}
  \label{fig:relic75}
\end{figure}

In Fig.~\ref{fig:relic75} the contours of preferred thermal relic
density in the representation \textbf{75} are plotted for one set of
(GUT scale) parameters, for which appreciable region for dark matter
within the WMAP limits is found, and which are not forbidden by the
$b\rightarrow s\gamma $ constraints.
Since the higgsino component in the representation \textbf{75} is
large \cite{Huitu:2005wh}, the resulting relic density is low, and
most of the parameter space is not overclosed by the WMAP limits.
Also the co-annihilations with the lightest chargino reduces the relic
density, since the lightest neutralino and chargino are nearly mass
degenerate in the higher $M_2$ part of the parameter space.
This is also seen in the Fig.~\ref{fig:relic75} at high $m_0$, where
the lightest chargino becomes the LSP for specific $M_2$ values.
In the low $M_2$ region the EWSB condition pushes the $\mu$ high,
which in turn decreases the higgsino component in the lightest
neutralino making it mostly a bino. The lightest neutralino and
chargino are not degenerate anymore, and the relic density increases
in low $M_2$ area, which enables the emergence of the WMAP preferred
region in the parameter space. The second lightest neutralino can
annihilate also directly into gauge bosons in this parameter region.
At the low $m_0$ region the LSP can be the lighter stop.
Again, increasing $\tan \beta$ enhances the higgsino component leading
to lower relic densities in general.

\begin{figure*}
\psfrag{m2}{$M_{2}$}
\psfrag{m0}{$m_{0}$}
 \centering 
 \subfloat[$\tan\beta = 10,\ \rm{sgn}(\mu) = +1,\ A_0=0$]{
   \label{fig:rd1-a}
   \includegraphics[width=0.46\textwidth]{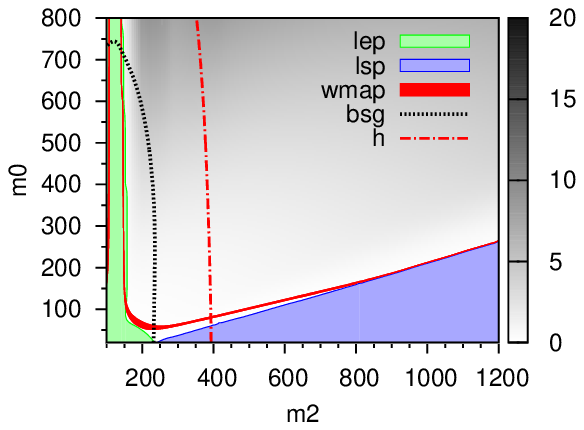}}
 \hspace{3mm} 
 \subfloat[$\tan\beta = 35,\ \rm{sgn}(\mu) = +1,\ A_0=0$]{
   \label{fig:rd1-b}
   \includegraphics[width=0.46\textwidth]{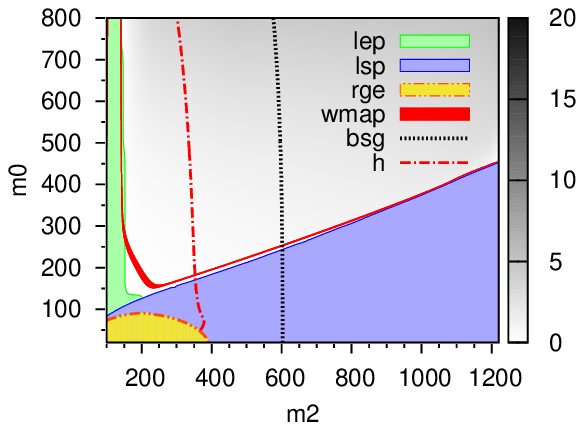}}\\
 \subfloat[$\tan\beta = 10,\ \rm{sgn}(\mu) = -1,\ A_0 = 0$]{
   \label{fig:rd1-c}
   \includegraphics[width=0.46\textwidth]{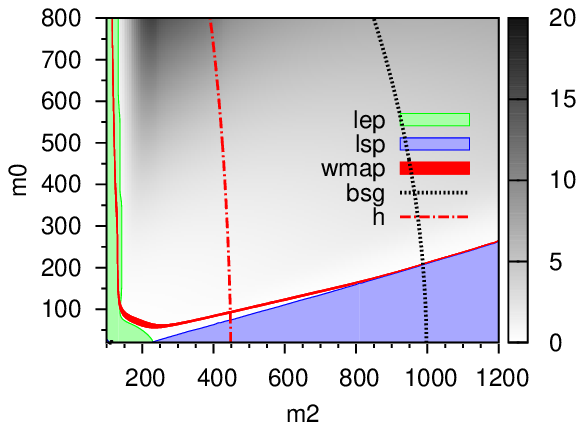}}
 \hspace{3mm} 
 \subfloat[$\tan\beta = 10,\ \rm{sgn}(\mu) = -1,\ A_0 = 1$ TeV]{
   \label{fig:rd1-d}
   \includegraphics[width=0.46\textwidth]{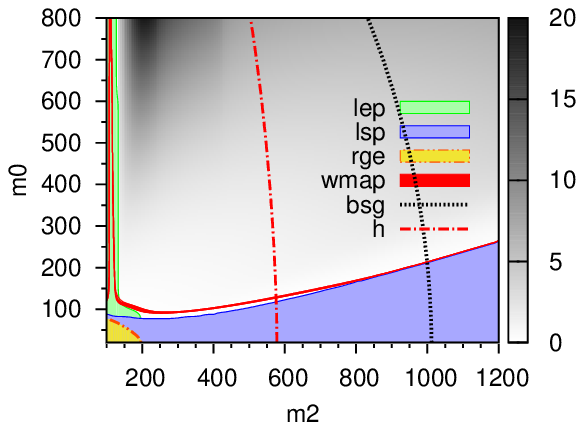}}
 \caption{Relic density $\Omega_\chi h^2$ in the representation
   \textbf{1} in ($M_{2}, m_0$) plane. 
    Explanation for the key is as in Fig.~\ref{fig:relic24}.
  }
 \label{fig:relic1}
\end{figure*}

\pagebreak

In Fig.~\ref{fig:relic1} the contours of preferred thermal relic
density in the representation \textbf{1} are plotted for four
different sets of (GUT scale) parameters for the reference.
For the chosen parameters the WMAP preferred regions are found near
the $M_2$ (\emph{i.e.} $m_{1/2}$ in mSUGRA and CMSSM language) and
$m_0$ axes.\footnote{ The hyperbolic branch/focus point region
  \cite{Chan:1997bi,Feng:1999mn}, which requires a substantially
  larger $m_0$, is not found for this setup (moreover, the low scale
  parameterization might be more relevant in describing this region
  anyway \cite{Baer:2005ky}). Also, the $A$-annihilation funnel is
  absent, since it requires larger $\tan\beta \sim 45-60$.}  The $b
\to s \gamma$ limits exclude large area in the figures \ref{fig:rd1-c}
and \ref{fig:rd1-d}, where the sign of the $\mu$-parameter is
negative.

In the representation \textbf{200} the higgsino mixing is large
\cite{Huitu:2005wh} and importantly, the bino-component is very small.
Also the co-annihilations with the lightest chargino, which is nearly
mass degenerate with the lightest neutralino, reduce the relic density
substantially, and the resulting relic density is tiny.  In contrast
to the \textbf{75} dimensional case, the $\mu$ parameter decreases
with decreasing $M_2$, so the bino component does not get very large.
The $b \to s \gamma$ constraint is largely within acceptable limits
(for negative $\mu$ a bit larger $M_2$ and $m_0$ are required than for
positive $\mu$) and the LSP is mostly the lightest
neutralino. Typically the Higgs mass is well above 91 GeV limit for
$\tan\beta \gsim 10$, and the 114 GeV line is around $M_2 \sim 600$
GeV. For the lower values of $\tan\beta$, the $M_2$ parameter must be
above 1 TeV.
The parameter space suitable for finding partial neutralino dark
matter can be extended both in $m_0$ and in $M_2$ beyond 1 TeV for
both signs of $\mu$, but neutralino in this representation can never
be the only source of dark matter.

\begin{figure*}
\psfrag{m2}{$M_{2}$}
\psfrag{m0}{$m_{0}$}
 \centering 
 \subfloat[$\tan\beta = 10,\ \rm{sgn}(\mu) = +1,\ A_0=0, M_3=700$ GeV]{
   \label{fig:rd1200-a}
   \includegraphics[width=0.46\textwidth]{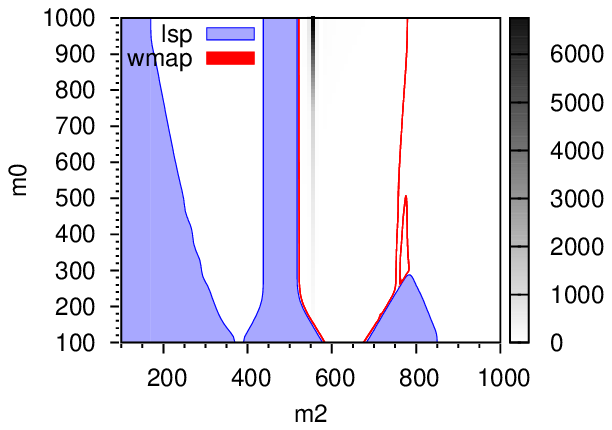}}
 \hspace{3mm} 
 \subfloat[$\tan\beta = 10,\ \rm{sgn}(\mu) = +1,\ A_0=0, M_3=400$ GeV]{
   \label{fig:rd1200-b}
   \includegraphics[width=0.46\textwidth]{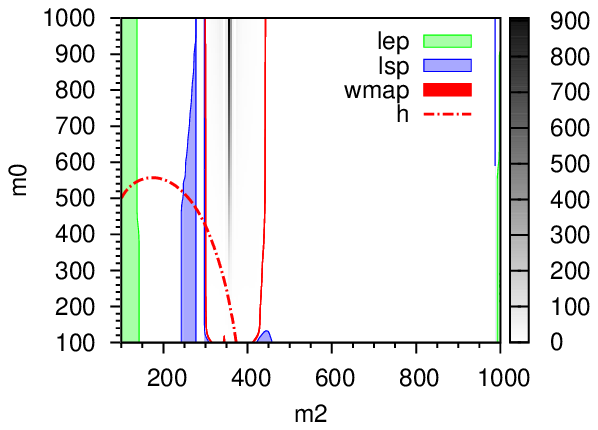}}
 \caption{Relic density $\Omega_\chi h^2$ in the representation
   $\mathbf{1} \oplus \mathbf{200}$ in ($M_{2}, m_0$) plane. 
    Red filling is the WMAP preferred region,
   \textsf{lep} shows an area next to the axes, where the experimental
   mass limits are not met and \textsf{lsp} the area where neutralino
   is not the LSP.  \textsf{h} gives the line of $m_h = 114$ GeV.}
 \label{fig:relic1200}
\end{figure*}

It is possible also to have many representations simultaneously.
For example, the Fig.~\ref{fig:relic1200} shows the relic density when
two representations, \textbf{1} and \textbf{200}, are combined. If one
allows the gaugino masses to be formed from two representations
simultaneously, then two of the gaugino masses can be chosen freely,
and only one is fixed by the representations. In
Fig.~\ref{fig:relic1200} the parameter $M_3$ is chosen to have a
specific value ($M_3=700$ GeV in Fig.~\ref{fig:rd1200-a} and $M_3=400$
GeV in Fig.~\ref{fig:rd1200-b}) and $M_2$ is allowed to vary.
The overall relic density is very low, as implied by the
representation \textbf{200}. However, there is a narrow peak where the
relic density is enormous. This is the area, where the gaugino masses
combine in such a way that the lightest neutralino is extremely
light. Therefore the annihilations are very much suppressed leading to
high relic density.
The position of the area where this happens depends on the
representations: e.g.~for the representation $\mathbf{1} \oplus
\mathbf{200}$ the parameter $M_1$ can be expressed as
\begin{equation}
  \label{eq:M1}
  M_1 = \frac{M_2 (1-10) + M_3 (10-2)}{1-2}  = 9 M_2 - 8 M_3,
\end{equation}
where the numbers come from the representation invariants. Therefore
the minimum of the lightest neutralino mass is positioned around
$M_2/M_3 \sim 8/9$.
The  $b \to s \gamma$ constraint is  within acceptable limits and the
Higgs mass is above limits (in Fig.~\ref{fig:rd1200-b} $m_h>113.5$ GeV 
in otherwise allowed regions).
By increasing the value of the $M_3$ mass parameter one effectively
increases the stop masses (due to the RG evolution), which then
increases the lightest Higgs mass. For example, the $m_h>117$ GeV
always in the Fig.~\ref{fig:rd1200-a}.
Chargino being the LSP excludes large areas in
Fig.~\ref{fig:rd1200-a}. The smaller area around $M_2\sim 800$ GeV at
the low $m_0$ region is due to the stau LSP. The $A$-threshold is also
shown as a small ring of higher relic density around $M_2\sim 770$
GeV.

Because with two representations there are two free gaugino masses,
the light LSP region is always present. Therefore, if the relic
density is low in general, there must also be a stripe of WMAP
preferred region available.
Changing the $M_3$ mass value the WMAP preferred stripe can be moved
through the parameter space, thus providing a method to fulfill the
WMAP constraints.

\section{Higgs in the neutralino cascade and dark matter}

\begin{figure*}
\psfrag{h}{\hspace{-1cm}$\sigma( \to h\tilde{\chi}_1^0 \to b\bar b)$ [pb]}
\psfrag{H}{\hspace{-1cm}$\sigma( \to H\tilde{\chi}_1^0  \to b\bar b)$ [pb]}
\psfrag{A}{\hspace{-1cm}$\sigma( \to A\tilde{\chi}_1^0  \to b\bar b)$ [pb]}
\psfrag{mg}{$m_{\tilde g}$}
\psfrag{ma}{$m_{A}$}
\psfrag{mA}{$m_{A}$}
\psfrag{tanb}{$\tan\beta$}
 \centering
 \subfloat[chain via $h$]{
   \label{fig:cs24-a}
   \includegraphics[width=0.45\textwidth,height=60mm]{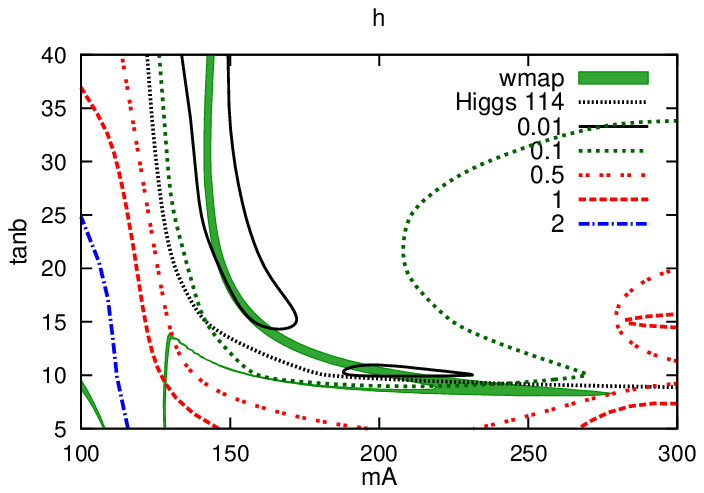}}
 \subfloat[chain via $h$]{
   \label{fig:cs24-b}
   \includegraphics[width=0.45\textwidth,height=60mm]{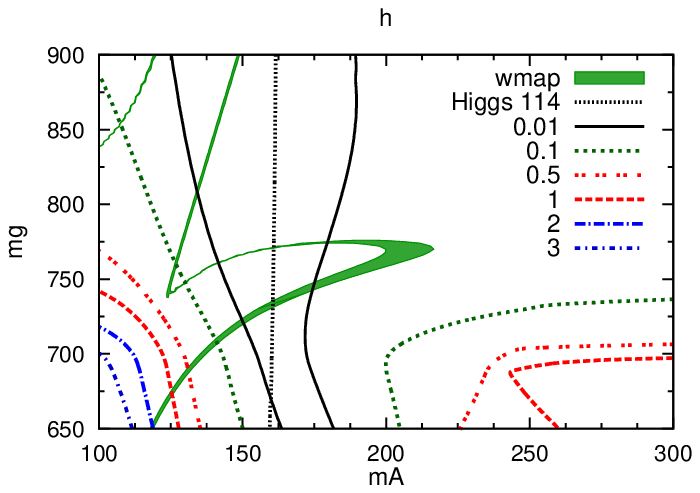}}\\
 \subfloat[chain via $H$]{
   \label{fig:cs24-c}
   \includegraphics[width=0.45\textwidth,height=60mm]{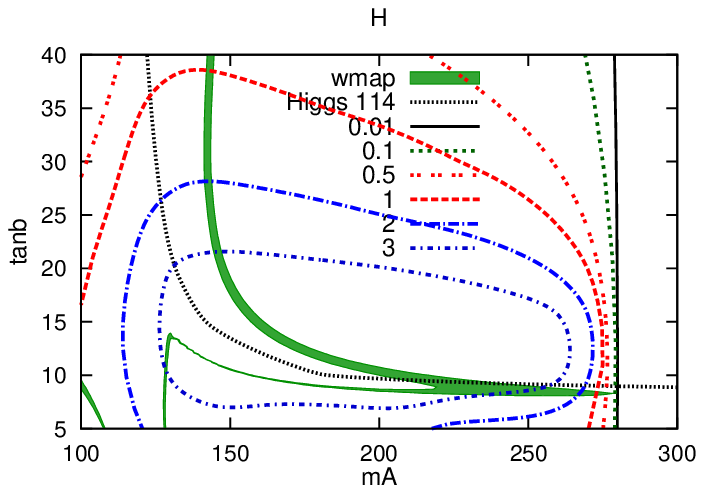}}
 \subfloat[chain via $H$]{
   \label{fig:cs24-d}
   \includegraphics[width=0.45\textwidth,height=60mm]{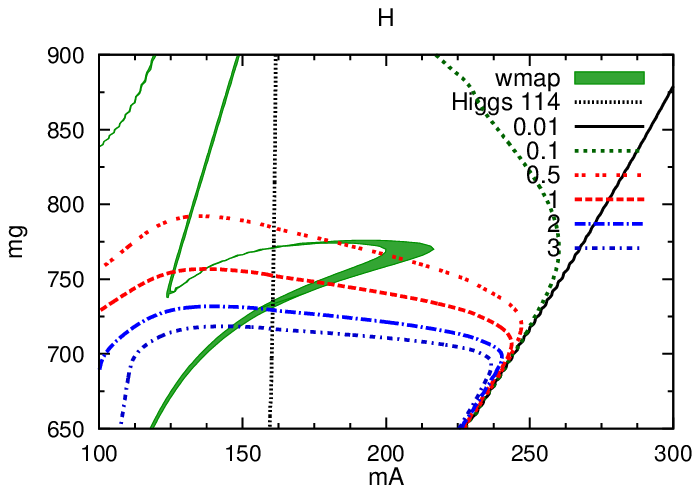}}\\
 \subfloat[chain via $A$]{
   \label{fig:cs24-e}
   \includegraphics[width=0.45\textwidth,height=60mm]{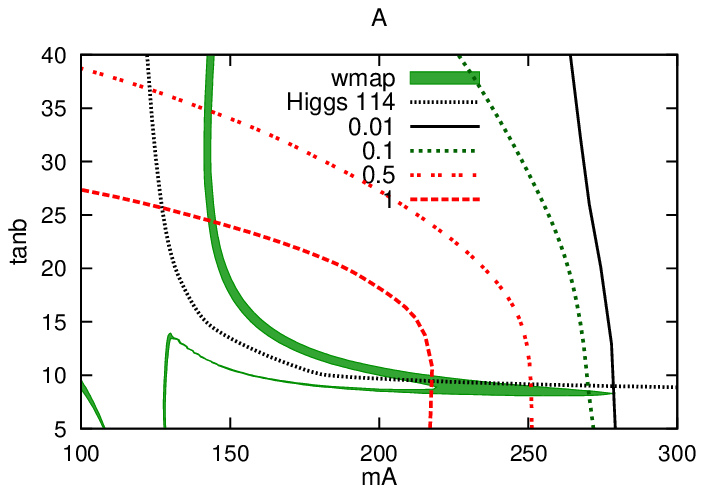}}
 \subfloat[chain via $A$]{
   \label{fig:cs24-f}
   \includegraphics[width=0.45\textwidth,height=60mm]{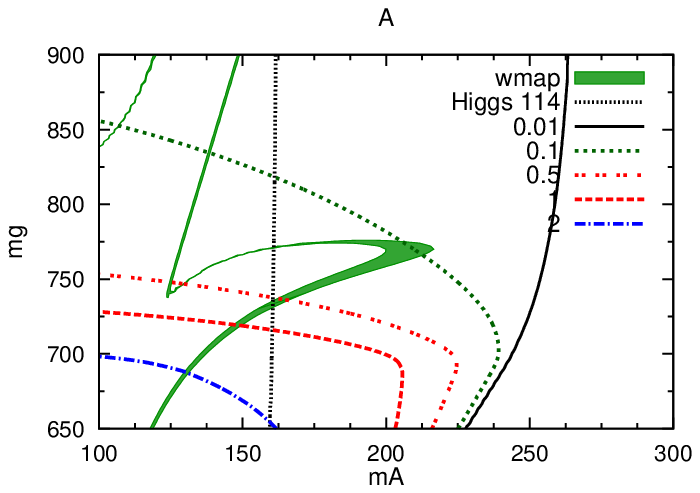}}
 \caption{Contours of constant cross section for $pp \to h/H/A +
   \tilde{\chi}_1^0 +X\to b \bar b + \tilde{\chi}_1^0 +X$ in the
   $(m_A, \tan\beta$) (left)
   and $(m_A, m_{\tilde g}$) (right) planes. The solid fill denotes
   the WMAP preferred relic density region and the dense dashed curve shows
   the Higgs 114 GeV mass limit (higher $m_A$ values correspond to a
   higher $m_h$).}
\label{fig:cs24}
\end{figure*}

Having discussed the basic features of the SU(5) non-universal gaugino
mass scenario, and the region of parameter space preferred by the
thermal relic density, we describe the numerical features of the cross
section of the signal events in the representation \textbf{24}. In
Fig.~\ref{fig:cs24} the cross section for $\tilde{g} \ra \tilde{q}q$,
$\tilde{q} \ra \tilde{\chi_2}q$, $ \tilde{\chi_2} \ra \tilde{\chi_1}
h/H/A$, $h/H/A \ra b\bar{b}$ produced in the pp interactions in the
$(m_A, \tan\beta)$ (figures
\ref{fig:cs24-a},~\ref{fig:cs24-c},~\ref{fig:cs24-e}) and $(m_A,
m_{\tilde g})$ (figures
\ref{fig:cs24-b},~\ref{fig:cs24-d},~\ref{fig:cs24-f}) planes are
plotted.  The solid (green) fill denotes the WMAP preferred relic
density region.  Also the Higgs 114 GeV mass contour is plotted with
the contours of constant cross section (larger $m_A$ values correspond
to larger $m_h$).
The values of the parameters are
($\mu$~=~+700~GeV/$c^2$, $m_{\tilde{\rm q}}$~=~600~GeV/$c^2$,
$m_{\tilde{\rm \ell}}$~=~350~GeV/$c^2$, and the trilinear coupling for
the top sector is chosen to be $A_t = 800$ GeV). In the $(m_A,
\tan\beta)$ figures the gluino mass is chosen as $m_{\tilde{\rm
    g}}$~=~770~GeV/$c^2$, and in the $(m_A, m_{\tilde g})$ plane
$\tan\beta = 10$.
In the $(m_A, \tan\beta)$ plane the cross section is largest for
the $H$ Higgs production. Also the stripe of WMAP preferred relic density
passes through the large cross section area. For the $A$ production, the
cross section is somewhat smaller, while for the $h$ production it is
substantially smaller for the WMAP preferred area.
The near-horizontal WMAP stripe around $\tan\beta = 8$ (beginning at
$m_A = 130$ GeV) corresponds to
the  closing of the Higgs resonance in the LSP
annihilation: in the region below that line the lightest neutralino mass is
more than half of the light Higgs boson mass, 
and thus the annihilation never occurs at the resonance.
Heavier $A$ correspond to the larger relic
density. The relic density is lowest in between the two relic
density stripes, the minimum occurring at the $A$-peak.

In the $(m_A, m_{\tilde g})$ plane the large $H$ production cross
section is again passed by the preferred relic density stripe.
The Higgs 114 GeV limit divides the
parameter space in two (vertical line in the figures
\ref{fig:cs24-b},~\ref{fig:cs24-d},~\ref{fig:cs24-f}; it should be
noted, though, that for the $\tan\beta = 10$ used here the actual
Higgs boson mass limit can be as low as around 90 GeV). 
Also here the
relic density is the lowest in between the two relic density stripes.
The horizontal kink in the relic density stripe around $m_{\tilde g}
\sim 770$ GeV corresponds to the closing of the Higgs resonance in the
LSP annihilation: in contrast to the  $(m_A, \tan\beta)$-plots, the Higgs
resonance is open below that line due to the decreasing of the
lightest neutralino mass.
The  $A$ and $h$ cross sections  follow the pattern of the  $(m_A,
\tan\beta)$ plane,  the $A$ production cross section being slightly 
smaller than for $H$ while
$h$ production is the one with the lowest cross section.
The cross sections and relic densities were calculated with {\tt PROSPINO}
\cite{Beenakker:1996ch}, {\tt SDECAY} \cite{Muhlleitner:2003vg},
 {\tt HDECAY} \cite{Djouadi:1997yw} and {\tt SuSpect}
\cite{Djouadi:2002ze} and {\tt micrOMEGAs} \cite{Belanger:2006is,Belanger:2004yn,Belanger:2001fz}.

Table \ref{tab:mass} shows the mass values for $\tilde{\chi_1}$ and
$\tilde{\chi_2}$ and the branching fractions for the $\tilde{\chi}_{2}
\ra \rm h/H/A\tilde{\chi}_{1}$ decay channels in the representations
$\mathbf{1}$ and $\mathbf{24}$ with m$_{\rm A}$~=~190~GeV/$c^2$ and in
the representation $\mathbf{24}$ with m$_{\rm A}$~=~210~GeV/$c^2$ with
tan$\beta$~=~10.  For the {\bf 24}-dimensional representation all
Higgs boson channels are possible. The branching fraction is largest
for the CP-even neutral Higgs boson H. As the gluino mass increases,
the branching fraction to $\tilde{\chi}_{2} \ra \rm h/H/A \ra
\tilde{\chi}_{1}$ decreases.  The signal cross sections are obtained
applying the $\tilde{\chi}_{2} \ra \rm h/H/A\tilde{\chi}_{1}$ and $\rm
h/H/A \ra \rm b\overline{\rm b}$ branching fractions on the total SUSY
production cross section in the chosen parameter point.  In the
singlet case, the $m_{\tilde g}=770$ GeV is too small to allow
kinematically the cascade decay.  We take $m_{\tilde g}=850$ GeV, with
which the decay through the light Higgs boson h is possible.

\begin{table*}
\small
\centering
\vskip 0.1 in
\caption{Mass values for $\tilde{\chi_1}$ and $\tilde{\chi_2}$ and the
  branching fractions for the $\tilde{\chi}_{2} \ra \rm
  h/H/A\tilde{\chi}_{1}$ decay channels in the representations
  $\mathbf{1}$ ($m_{\tilde g} =850$ GeV) and $\mathbf{24}$ ($m_{\tilde g}
  =770$ GeV) with m$_{\rm A}$~=~190~GeV/$c^2$ and in the
  representation $\mathbf{24}$ with m$_{\rm A}$~=~210~GeV/$c^2$ and
  tan$\beta$~=~10.}
\begin{tabular}{|l|c|c|c|c|c|}
  \hline
  Representation, $\rm m_{\rm A}$  &  $\rm m_{\tilde{\chi_1}}$ (GeV/$c^2$)
&  $\rm m_{\tilde{\chi_2}}$ (GeV/$c^2$) & BR($\tilde{\chi}_{2} \ra \rm h\tilde{\chi}_{1}$)
&  BR($\tilde{\chi}_{2} \ra \rm H\tilde{\chi}_{1})$ &  BR($\tilde{\chi}_{2} \ra \rm A\tilde{\chi}_{1})$ \\
  \hline
  $\mathbf{1}$, 190~GeV/c$^2$  & 120.6 & 246.8 & 0.897 & 0 & 0\\
  $\mathbf{24}$, 190~GeV/c$^2$ & 50.5 & 307.4 & 0.00478 & 0.718 & 0.205\\
  $\mathbf{24}$, 210~GeV/c$^2$ & 50.5 & 307.5 & 0.00144 & 0.739 & 0.161\\
  \hline
\end{tabular}
\label{tab:mass}
\end{table*}

\section{Event generation}

\subsection{SUSY cascades}

The event generation and simulation were performed and the results were made 
public in the CMS framework \cite{cmsweb}.
The Higgs boson production in SUSY cascades was generated with PYTHIA 
\cite{pythia} using a squark and gluino production with general MSSM simulation.
The CTEQ5L structure functions \cite{cteq5} were used. The simulated point was $\mu$~=~+700~GeV/$c^2$, 
$m_{\tilde{\rm q}}$~=~600~GeV/$c^2$,
$m_{\tilde{\rm \ell}}$~=~350~GeV/$c^2$,
and the trilinear coupling for the top sector was chosen to be $A_t = 800$ GeV. 
The gluino mass was chosen as $m_{\tilde{\rm g}}$~=~770~GeV/$c^2$.
The U(1), SU(2) and SU(3) gaugino masses were chosen separately for each scenario,
values shown in Table \ref{table:gauginoMasses}. PYTHIA version 6.325 was used
within the CMKIN framework \cite{cmsim}. No forcing of decay channels
was used. After the event generation, the events were sorted by the Higgs boson content,
giving a sample of signal events, and a sample of SUSY background events 
with no Higgs bosons in the event. 
The SUSY production of multi-jet events is expected to be the main source of background.

In the Singlet representation a cross section of 18.3 pb was used for the SUSY cascade 
(including Higgs boson production) with $\rm m_{\rm A}$~=~190~GeV/c$^2$, tan$\beta$ = 10.
The Higgs boson production cross section is 4.13~pb for this point.
In the {\bf 24}-dimensional representation with
$\rm m_{\rm A}$~=~190 GeV/c$^2$, tan$\beta$ = 10, the cross sections are 23.1 pb and 5.07 pb,
and with $\rm m_{\rm A}$~=~210 GeV/c$^2$, tan$\beta$ = 10, 23.1 pb and 4.97 pb, respectively.

\begin{table*}
\small
\centering
\vskip 0.1 in
\caption{Gaugino mass parameters for the studied points.}
\begin{tabular}{|l|c|c|c|}
  \hline
Representation, $\rm m_{\rm A}$ & U(1)(GeV/$c^2$) & SU(2)(GeV/$c^2$) & SU(3)(GeV/$c^2$) \\
  \hline
{\bf 1}, 190 GeV/$c^2$ & 122.9 & 245.3 & 850 \\
{\bf 24}, 190 GeV/$c^2$ & -44.1 & -264 & 610 \\
{\bf 24}, 210 GeV/$c^2$ & -45.54 & -272.7 & 630 \\
  \hline
\end{tabular}
\label{table:gauginoMasses}
\end{table*}

\subsection{Standard Model backgrounds}

The $\rm t\overline{\rm t}$, QCD multi-jet, Z+jets and W+jets production are
the main SM backgrounds for the Higgs boson searches with $\rm H/A/h \ra \rm b\bar{\rm b}$ in
SUSY cascades. The $\rm t\overline{\rm t}$ events were generated
with PYTHIA \cite{pythia}. The production cross section was normalized
to the NLO cross section of 840~pb. The QCD multi-jet background for the light 
Higgs boson h in the SUSY cascades has been studied  with full 
simulation in Ref.~\cite{CMSNote2006/090} and has been found to be 
negligible requiring energetic jets, two b jets and 
large missing $\rm E_{\rm T}$ ($>$~200~GeV). This background particularly sensitive to the
detector performance through the jet measurements. Therefore it was studied in this work
with the new full simulation \cite{cmssw} for the $\rm E_{\rm T}^{\rm miss}$ cut.
Events for the full simulation study were generated with PYTHIA in two bins: 120 $< \rm p_{\rm T} <$ 170 GeV/c and 
170~$<\rm p_{\rm T} <$~230~GeV/c.
 The Z+jets and W+jets events can contribute to the background through large missing $\rm E_{\rm T}$ from
  $\rm Z \ra \nu\nu$ decays and from lost leptons. 
These backgrounds have been investigated for the inclusive SUSY searches with ALPGEN \cite{alpgen}
and PYTHIA simulation.
The Z+jets and W+jets backgrounds have been found to be roughly on the same level
as the $\rm t\overline{\rm t}$ background after the basic 
selections for the inclusive search: $\rm E_{\rm T}^{\rm miss} >$~200~GeV and four jets 
with $\rm E_{\rm T} >$~ 30 GeV \cite{pTDR2}. The requirement of two b jets with 
$\rm E_{\rm T} >$~100~GeV
will suppress these backgrounds well below the  
$\rm t\overline{\rm t}$ background and were therefore not simulated in this work.

\section{Detector simulation}

The response of the CMS detector was simulated with the fast simulation
package FAMOS \cite{famos}.
The pile-up corresponding to the luminosity of 2$\times$10$^{33}\rm cm^{-2}\rm s^{-1}$
was included. 
The physics objects were reconstructed with the standard
methods available in the CMS reconstruction software.
The primary vertex was reconstructed and selected with an algorithm searching for 
the highest sum of the transverse momenta of the associated tracks \cite{orca}. 
Jets were reconstructed in a cone of 0.5. The b tagging was performed with a 
robust track counting based B-Tagging Algorithm of CMS \cite{note2006_014}.
The QCD MET reconstruction and the MET cut efficiency was studied in addition 
 with CMS full simulation \cite{cmssw}. 

\section{Event selection}

The signal events are characterized by many energetic hadronic jets.
Due to the  $m_{\tilde{\rm g}}> m_{\tilde{\rm q}}$ condition, the gluino decays to 
a quark and the corresponding squark. Therefore the cascade producing the Higgs boson
contains two energetic jets from the decays of the gluino and the squark and 
two b jets from the decay of the Higgs boson. At least two jets are produced in the 
opposite cascade. The two LSP's ($\tilde{\chi}_1^0$) at the
end of the two cascades lead to large missing $\rm E_{\rm T}$. These characteristics
can be used to suppress the SM backgrounds, for which the jet multiplicities and 
missing $\rm E_{\rm T}$ are typically significantly smaller.

\subsection{Trigger}

The events are supposed to be triggered with a multi-jet trigger at the Level-1 (L1) and with a combined
multi-jet plus missing  $\rm E_{\rm T}$ trigger at the High Level Trigger (HLT) \cite{L1_TDR,DAQ_TDR}. 
The $\rm E_{\rm T}$ thresholds for the four jets at the trigger level
were taken to be 80, 50, 30 and 10~GeV. The threshold for the missing  $\rm E_{\rm T}$
calculated from the calorimeter cells was fixed to 135~GeV. The Level-1 trigger was not simulated. 
The effect of this choice is negligible since
the efficiency of the multi-jet trigger on the SUSY events has been found to be close to 100\%
\cite{CMSNote2006/090}. In this work the HLT efficiency 
in the SUSY events is found to be 67\%. The  $\rm t\overline{\rm t}$ 
background is efficiently reduced already at the trigger level 
with the missing  $\rm E_{\rm T}$ cut. The HLT efficiency for this sample was found to be 3.8\%.

\subsection{Jet selection}

\begin{figure*}
  \centering
  \vskip 0.1 in
  \begin{tabular}{ccc}
  \begin{minipage}{5cm}
    \centering
    \resizebox{\linewidth}{60 mm}{\includegraphics{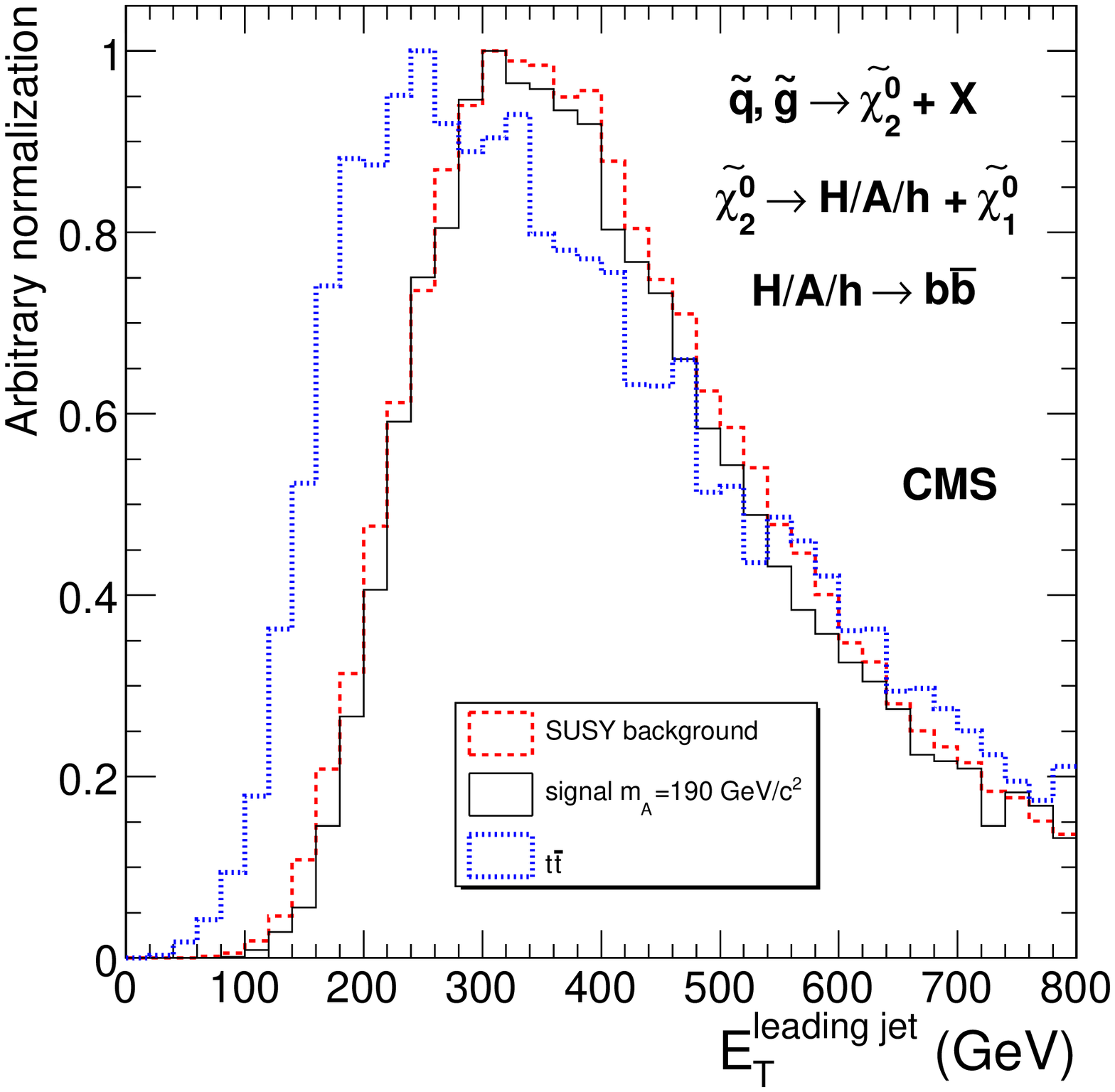}}
\put(-130,10){$a$}
  \end{minipage}
  &
  \begin{minipage}{5cm}
    \centering
    \resizebox{\linewidth}{60 mm}{\includegraphics{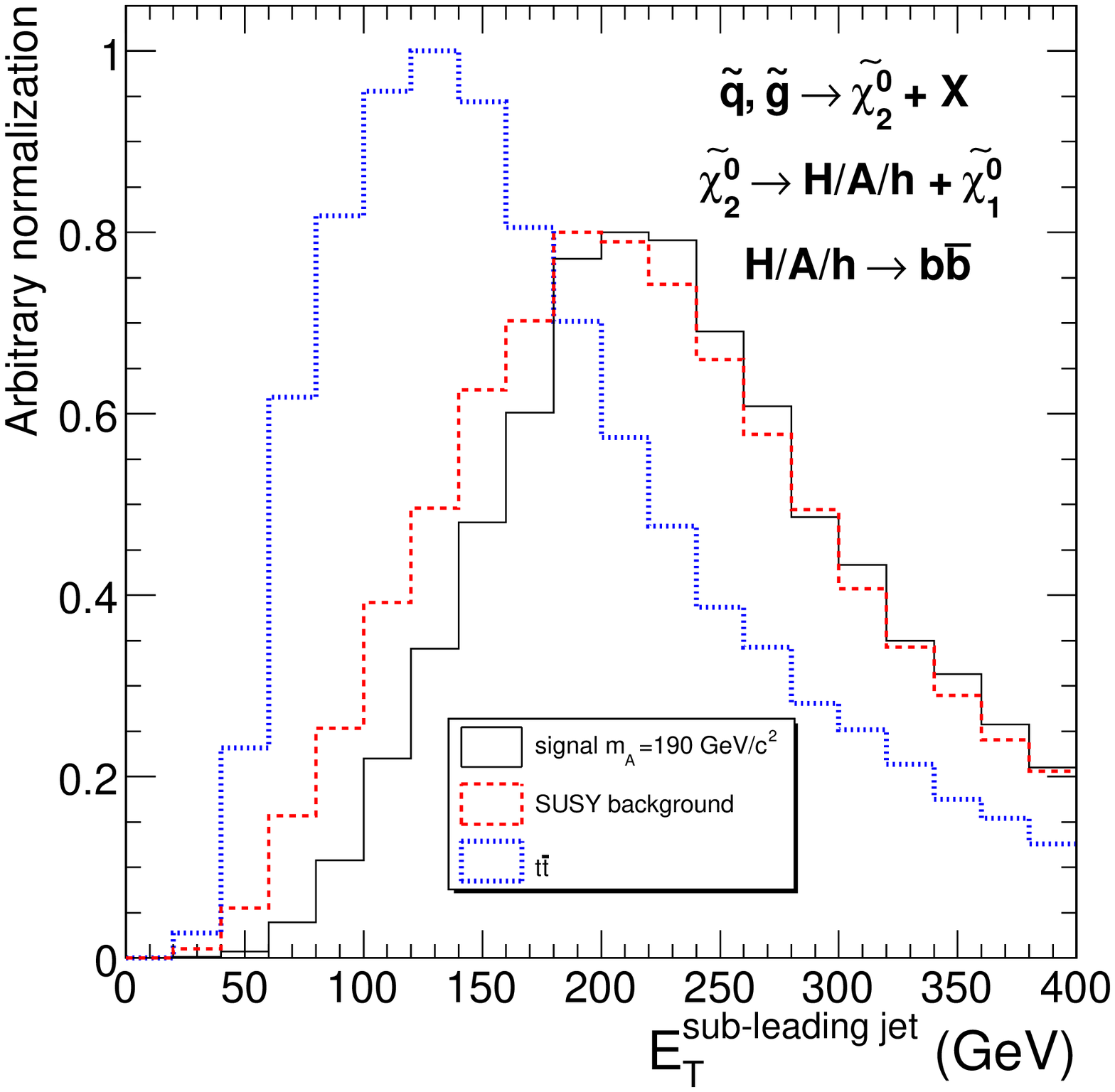}}
\put(-130,10){$b$}
  \end{minipage}
  &
  \begin{minipage}{5cm}
    \centering
    \resizebox{\linewidth}{60 mm}{\includegraphics{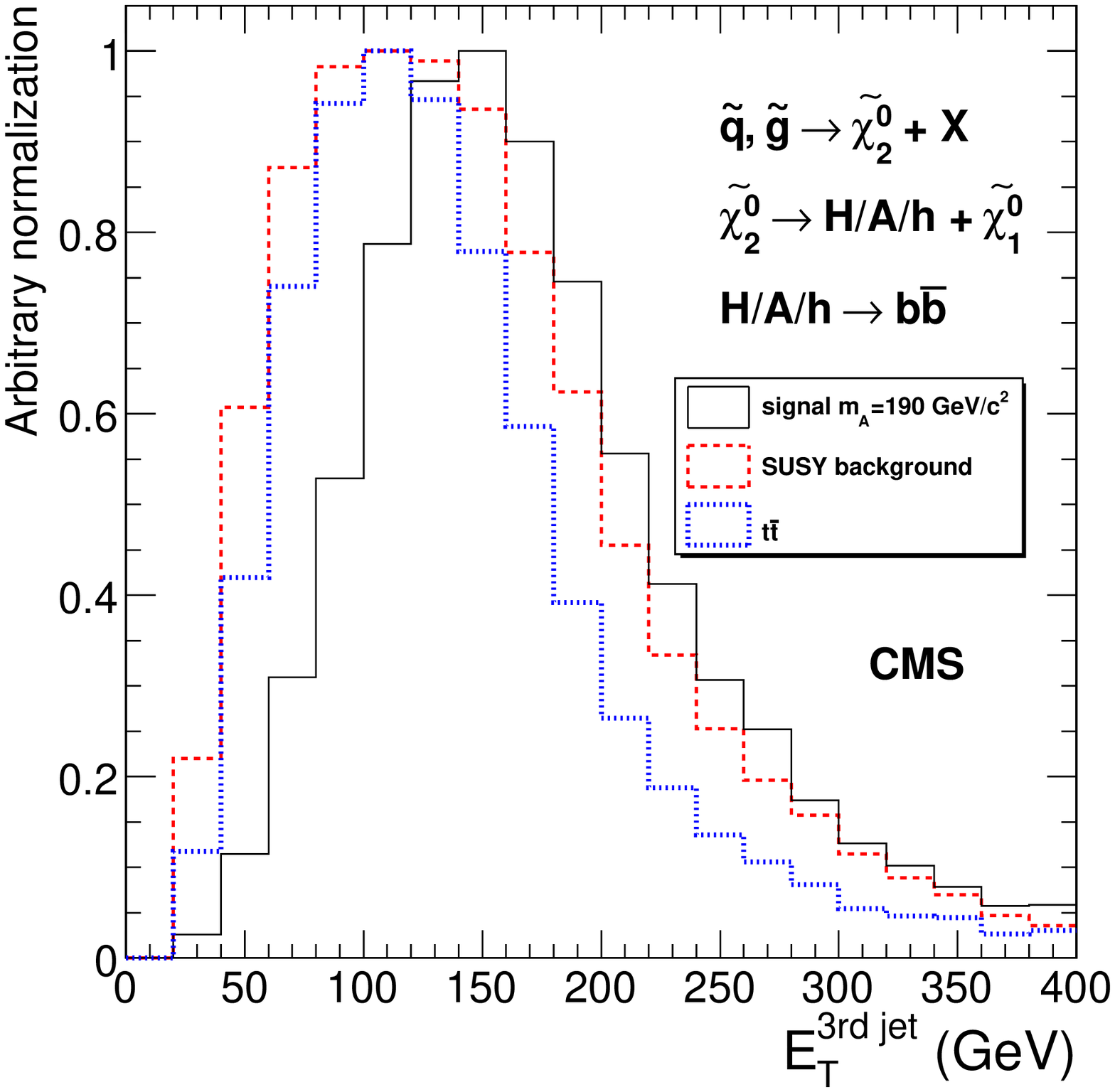}}
\put(-130,10){$c$}
  \end{minipage}
  \end{tabular}
  \caption{Distribution of $\rm E_{\rm T}$ of the most energetic jet ($a$), the sub-leading jet ($b$),
and the third most energetic jet ($c$) for the signal with m$_{\rm A}$~=~190~GeV/$c^2$
and tan$\beta$ = 10 (solid line), for the SUSY background (dashed line) and for the $\rm t\overline{\rm t}$
background (dotted line) in the representation {\bf 24}. }
    \label{fig:jets}
\end{figure*}

\begin{figure*}
  \centering
  \vskip 0.1 in
  \begin{tabular}{ccc}
  \begin{minipage}{5cm}
    \centering
    \resizebox{\linewidth}{60 mm}{\includegraphics{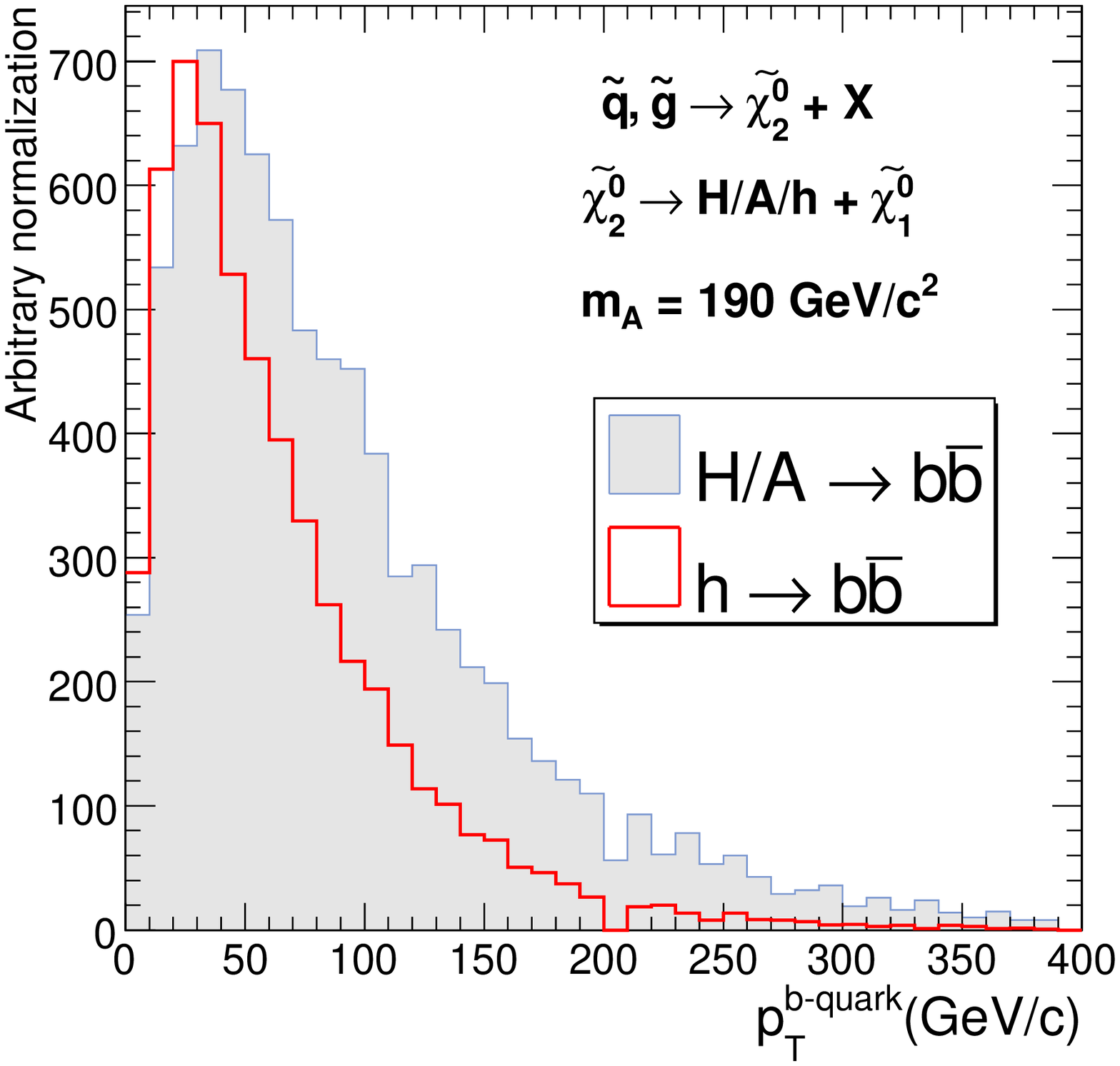}}
  \caption{Distribution of $\rm p_{\rm T}$ for the b quarks from
 $\rm H/A \ra \rm b\overline{\rm b}$ (filled histogram) and from
$\rm h \ra \rm b\overline{\rm b}$ (solid line) for the signal with m$_{\rm A}$~=~190~GeV/$c^2$ 
in the representation {\bf 24}.\newline}
  \label{fig:ptbquarks}
  \end{minipage}
  &
  \begin{minipage}{5cm}
    \centering
    \resizebox{\linewidth}{60 mm}{\includegraphics{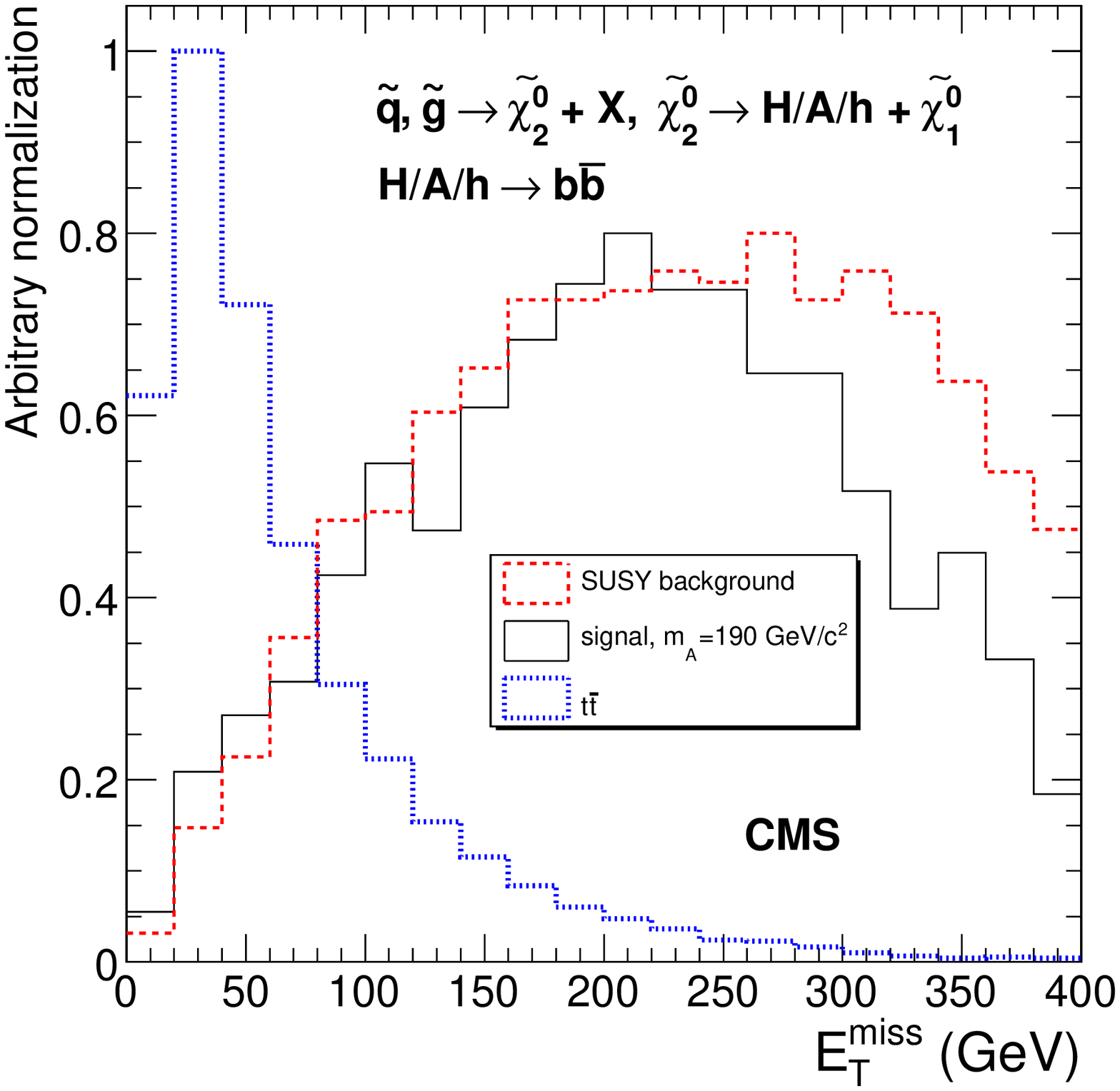}}
  \caption{Distribution of missing $\rm E_{\rm T}$
for the signal with m$_{\rm A}$~=~190~GeV/$c^2$
and tan$\beta$ = 10 (solid line) in the representation {\bf 24}, for the SUSY background (dashed line)
 and for the $\rm t\overline{\rm t}$
background (dotted line).}
  \label{fig:chi2hbb_EtMiss}
  \end{minipage}
  &
  \begin{minipage}{5cm}
    \centering
    \resizebox{\linewidth}{60 mm}{\includegraphics{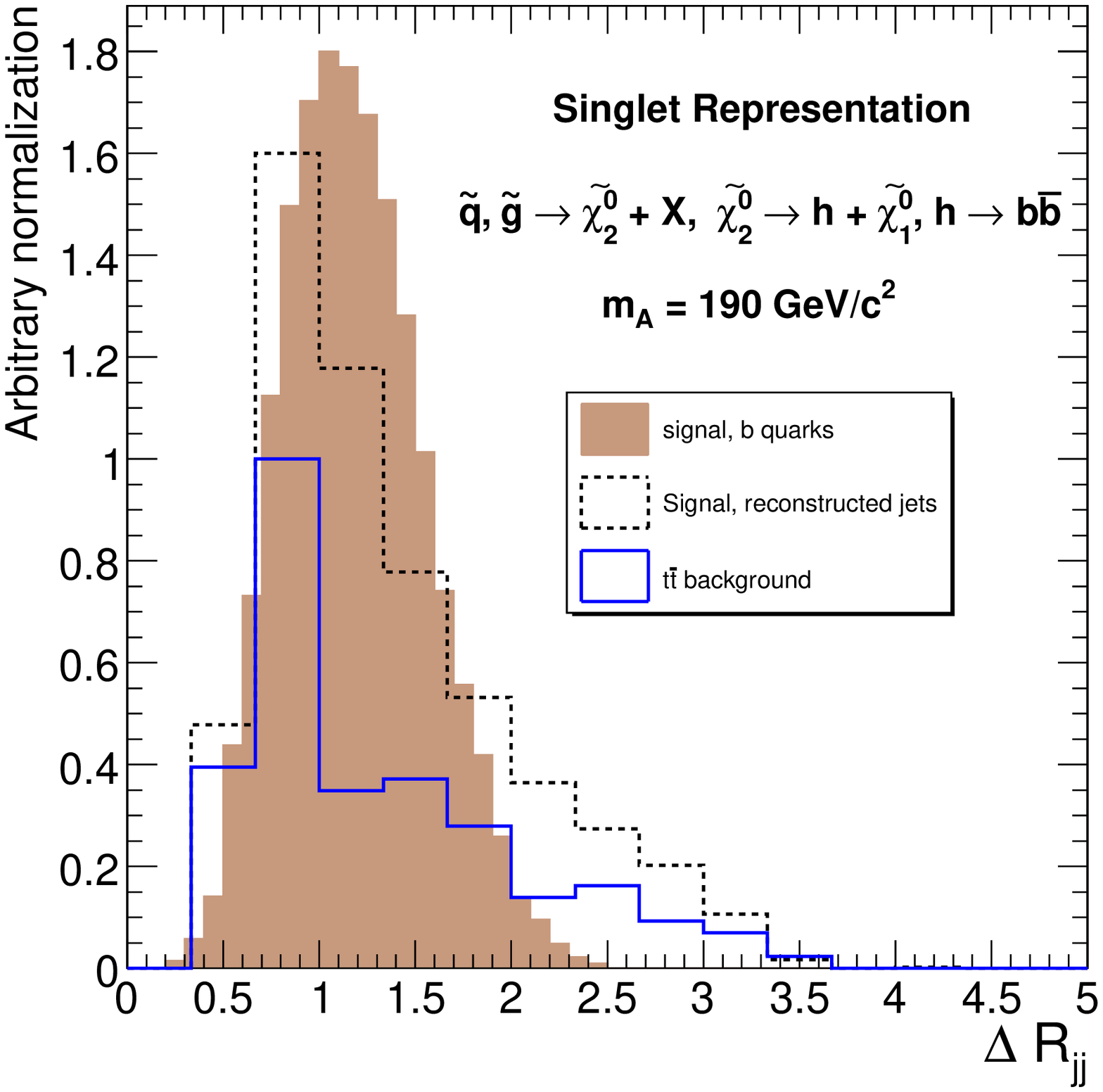}}
  \caption{Distribution of $\Delta$R between the selected jets for the signal
in the singlet representation with m$_{\rm A}$~=~190~GeV/$c^2$. and for the $\rm t\bar{\rm t}$
background. The $\Delta$R between the b quarks from the Higgs boson decay also shown. }
  \label{fig:drjj_rep1}
  \end{minipage}
  \end{tabular}
\end{figure*}

Figure \ref{fig:jets} shows the $\rm E_{\rm T}$
distributions for the three most energetic jets
in the signal, in the SUSY background and in the $\rm t\overline{\rm t}$ events.
For the off-line analysis, events with four jets with $\rm E_{\rm T}>$~50~GeV were first searched for.
At this level, the $\rm E_{\rm T}$ cuts were set to 250, 200, 100~GeV for the three most
energetic jets. These cuts are more stringent than those used in Ref.~\cite{CMSNote2006/090}
and thus can help to suppress more efficiently the QCD multi-jet and W+jets and Z+jets backgrounds.
The $\rm t\overline{\rm t}$ background is reduced to 35\% with the $\rm E_{\rm T}^{\rm jet}$ cuts.
Two b jets are expected from the Higgs boson decay. Figure \ref{fig:ptbquarks}
shows the  $\rm p_{\rm T}$ distributions
for the two b quarks from $\rm H/A \ra \rm b\overline{\rm b}$. Efficiency for tagging one selected
jet as a b jet in the signal events, with at least one Higgs boson decaying to b quarks, was found to be about 30\%.
In addition to the possibility of mistagging a hadronic jet in the SUSY background, there are several sources of genuine
b jets, like $\tilde{\chi}_{2} \ra \rm Z \tilde{\chi}_{1}, \rm Z \ra \rm b\overline{\rm b}$
 and gluon splitting processes in hadronic jets.
As a consequence, the tagging efficiency in the SUSY background was found to be substantial, about 13\%.

Figure \ref{fig:chi2hbb_EtMiss} shows the distribution of the missing $\rm E_{\rm T}$ for the signal,
for the SUSY background and for the $\rm t\overline{\rm t}$ events.
Large missing $\rm E_{\rm T}$ is typical for SUSY events
due to the presence of neutralinos in the final state.
The $\rm t\overline{\rm t}$ background
is further suppressed with this cut by a factor of $\sim$~5.
The QCD multi-jet production contributes to the background through missing $\rm E_{\rm T}$ due to detector
resolution and mis-measurement of jets. The 
missing $\rm E_{\rm T}$ selection was studied with the new full simulation package \cite{cmssw}. 
Efficiency for $\rm E_{\rm T}^{\rm miss}>$~200~GeV was found to be 4$\times$10$^{-5}$ 
for 120 $< \rm p_{\rm T} <$ 170 GeV/c and 
9$\times$10$^{-5}$ for 170~$<\rm p_{\rm T} <$~230~GeV/c.
No trigger simulation was included in the estimation of this background.

Due to the large jet multiplicity and the mistagging possibility,
the combinatorial background from false jet assignment
can be expected to be large for the reconstruction of the Higgs boson invariant mass from the b jets.
This combinatorial background can be significantly reduced with a method
called a hemisphere separation technique, devised in \cite{hemisphereMethod}.
This method separates the jets into two ``hemispheres'' where the
initial gluino or squark jet originates, starting from the most energetic
jet in the two hemispheres. The assignment of a jet to the axis is performed exploiting the angles and
invariant masses. The method has been shown to be
effective in removing the combinatorial background arising from jet multiplicity in
SUSY events in Ref.~\cite{CMSNote2006/090}.

After the separation to hemispheres the b jets were selected in each hemisphere separately.
Due to the large mass of the decaying neutralino,
the Higgs boson, in particular the light Higgs boson h, receives a large boost leading to
a small separation angle between the b jets. The $\Delta$R ($\Delta \rm R = \sqrt{(\Delta\eta)^2 +(\Delta \phi)^2}$)
between the two b quarks from the Higgs boson, shown in Fig.~\ref{fig:drjj_rep1} for the light Higgs boson,
is peaked around $\Delta \rm R \sim$1.
Therefore the pair of b jets with the lowest $\Delta$R value was selected in each hemisphere.
Among these two combinations the one with lower $\Delta$R value was taken to present the Higgs boson.

\begin{figure*}
  \centering
  \vskip 0.1 in
  \begin{tabular}{cc}
  \begin{minipage}{7.5cm}
    \centering
    \resizebox{\linewidth}{60 mm}{\includegraphics{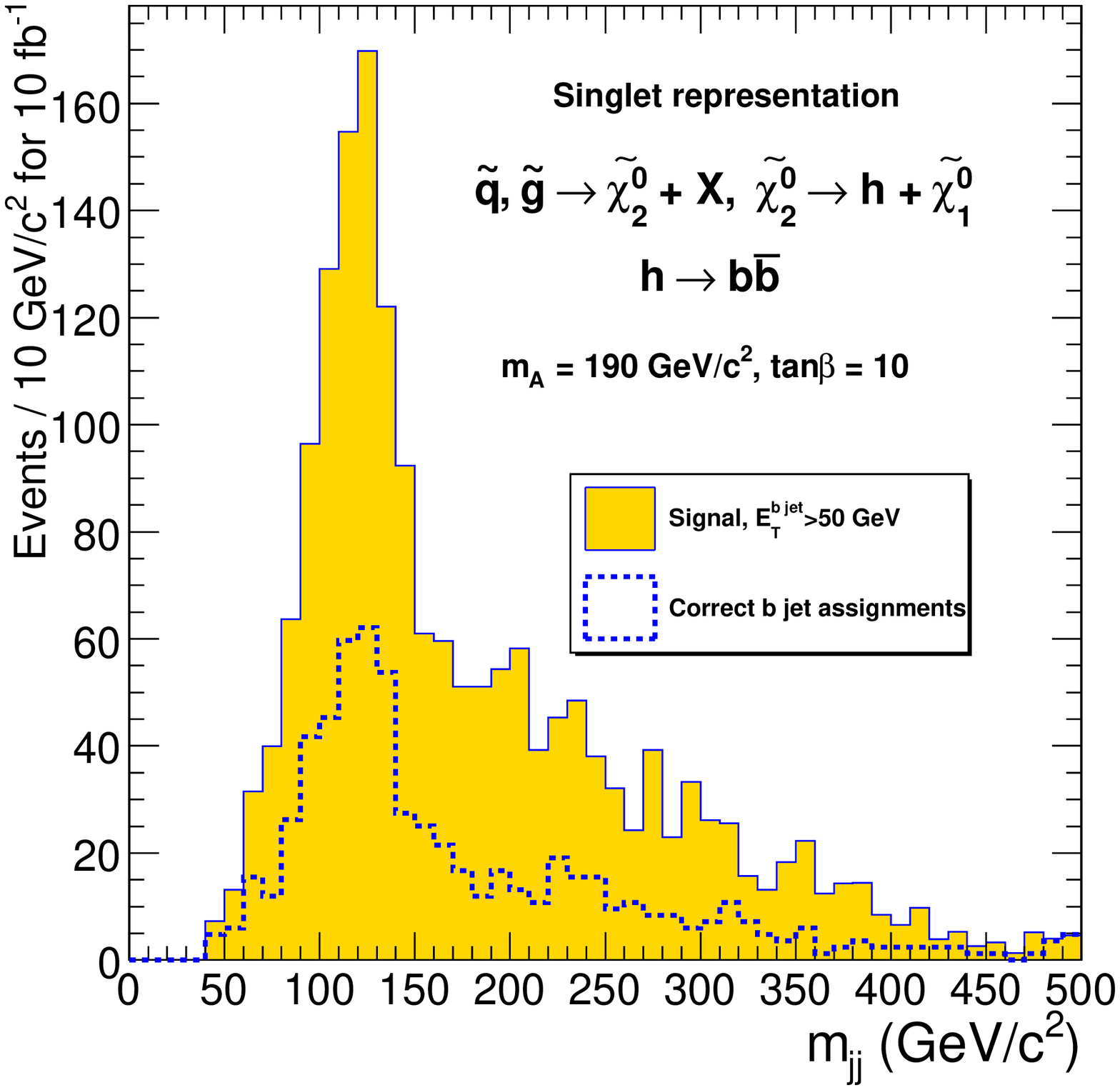}}
  \caption{Invariant mass of the two b jets (filled histogram) in the singlet representation. 
The mass distribution with correct b jet assignments is shown with dashed line.}
  \label{fig:hmassRealbb}
  \end{minipage}
  &
  \begin{minipage}{7.5cm}
    \centering
    \resizebox{\linewidth}{60 mm}{\includegraphics{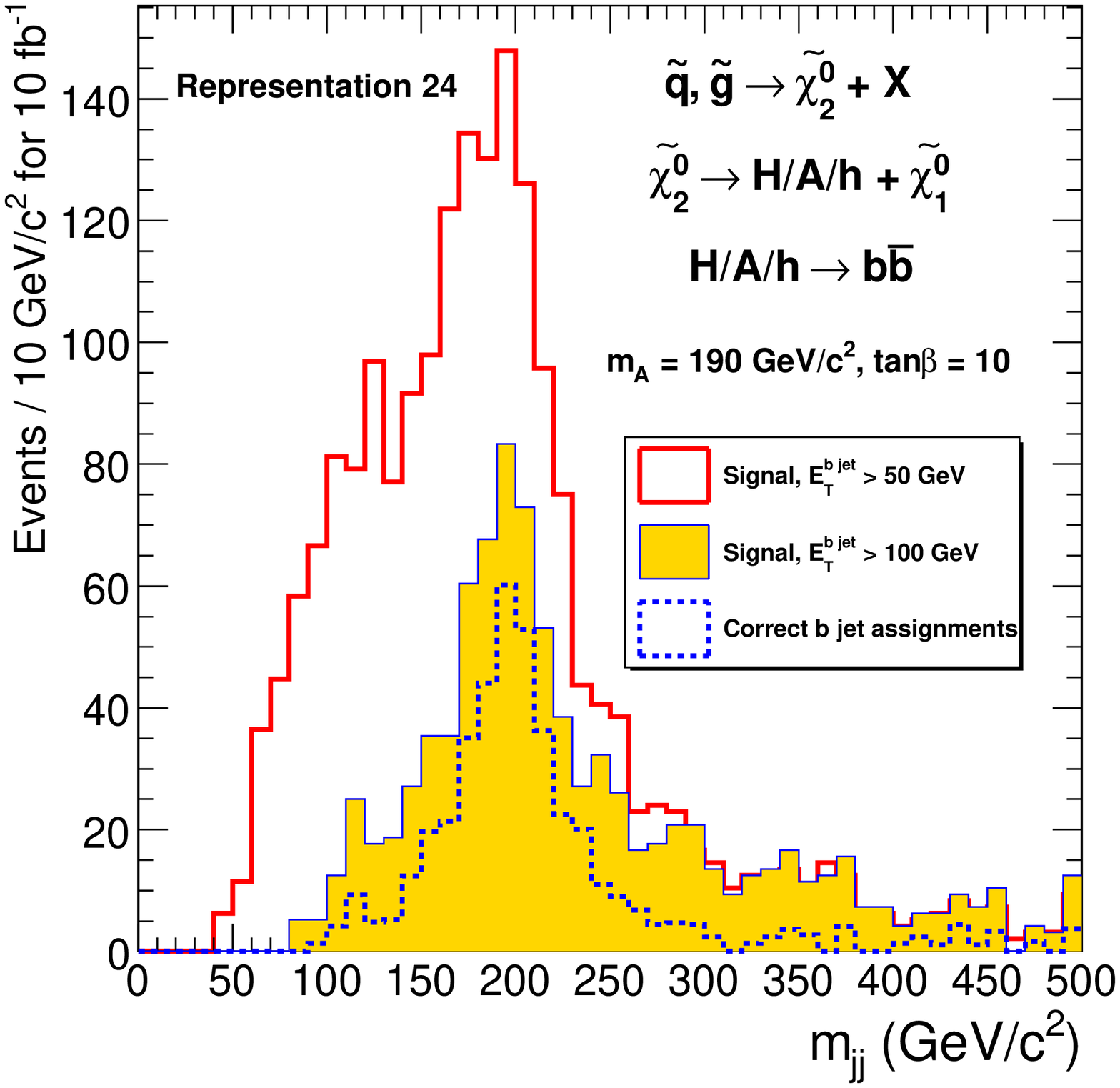}}
  \caption{As in Fig. \ref{fig:hmassRealbb}, for {\bf 24} representation with two 
different $\rm E_{\rm T}^{\rm b jet}$ cuts. The mass distribution with correct 
b jet assignments is shown for $\rm E_{\rm T}^{\rm b jet}>$~100~GeV.}
  \label{fig:HmassRealbb}
  \end{minipage}
  \end{tabular}
\end{figure*}

Figure \ref{fig:hmassRealbb} shows the $\rm b\overline{\rm b}$ mass distribution for the signal in the singlet
representation with m$_{\rm A}$~=~190~GeV/$c^2$. Only the light Higgs boson can be produced 
due to the small mass difference between the two neutralinos. The mass distribution with correct b-jet mass
assignments is also shown in the figure. The tagged b jet is defined as a genuine b jet if a b quark is
found within $\Delta R <$~0.4 from the jet axis. The fraction of correct jet assignments is 29\%. 
Large fraction of the false b-jet assignments is distributed to large mass values. A tail at large masses is 
visible also in the mass distribution with correct assignments. This tail is due to overlapping jets in the 
multi-jet environment. 
Figure \ref{fig:HmassRealbb} shows the $\rm b\overline{\rm b}$ mass distribution for the signal in the
representation {\bf 24} with m$_{\rm A}$~=~190~GeV/$c^2$. 
For this parameter point the mass difference between the neutralinos is large and the decay branching fractions
to heavy Higgs bosons are significant. The fraction of the light Higgs boson after all selection cuts is 3.5\%.
For the heavy Higgs bosons, 
better signal mass resolution and signal visibility was obtained with a significantly harder cut for the selected b jets,
 $\rm E_{\rm T}>$~100~GeV, as is shown in the Fig.~\ref{fig:HmassRealbb}.
About 87\% of the signal is due to the heavy CP-even Higgs boson H. The mass distribution with correct b-jet mass
assignments and with  $\rm E_{\rm T}>$~100~GeV is also shown in the figure. 
The opening angle in the Higgs boson decay is larger than for the light Higgs boson
h in the singlet representation making more favorable the jet reconstruction and b-jet identification.
The fraction of correct jet assignments was found to be 39\%.

\subsection{Signal over background ratios}

\begin{figure*}
  \centering
  \vskip 0.1 in
  \begin{tabular}{ccc}
  \begin{minipage}{7.5cm}
    \centering
  \resizebox{\linewidth}{60 mm}{\includegraphics{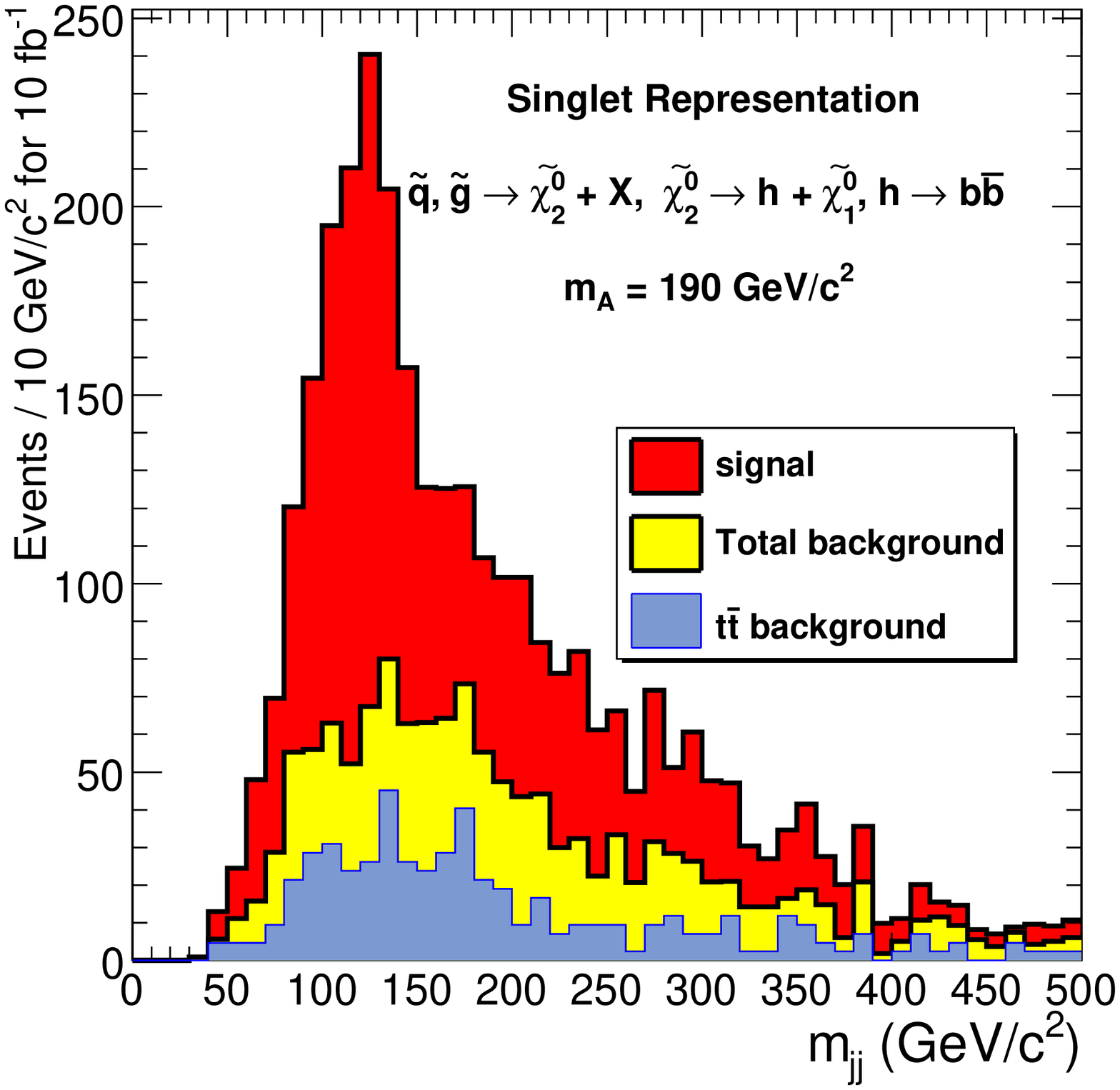}}
  \caption{Distribution of the invariant mass reconstructed from the selected b jets
for the signal, for the total background and for the $\rm t\overline{\rm t}$
background in the singlet representation with m$_{\rm A}$~=~190~GeV/$c^2$.}
  \label{fig:hmassRep1}
  \end{minipage}
  &
  \begin{minipage}{7.5cm}
    \centering
  \resizebox{\linewidth}{60 mm}{\includegraphics{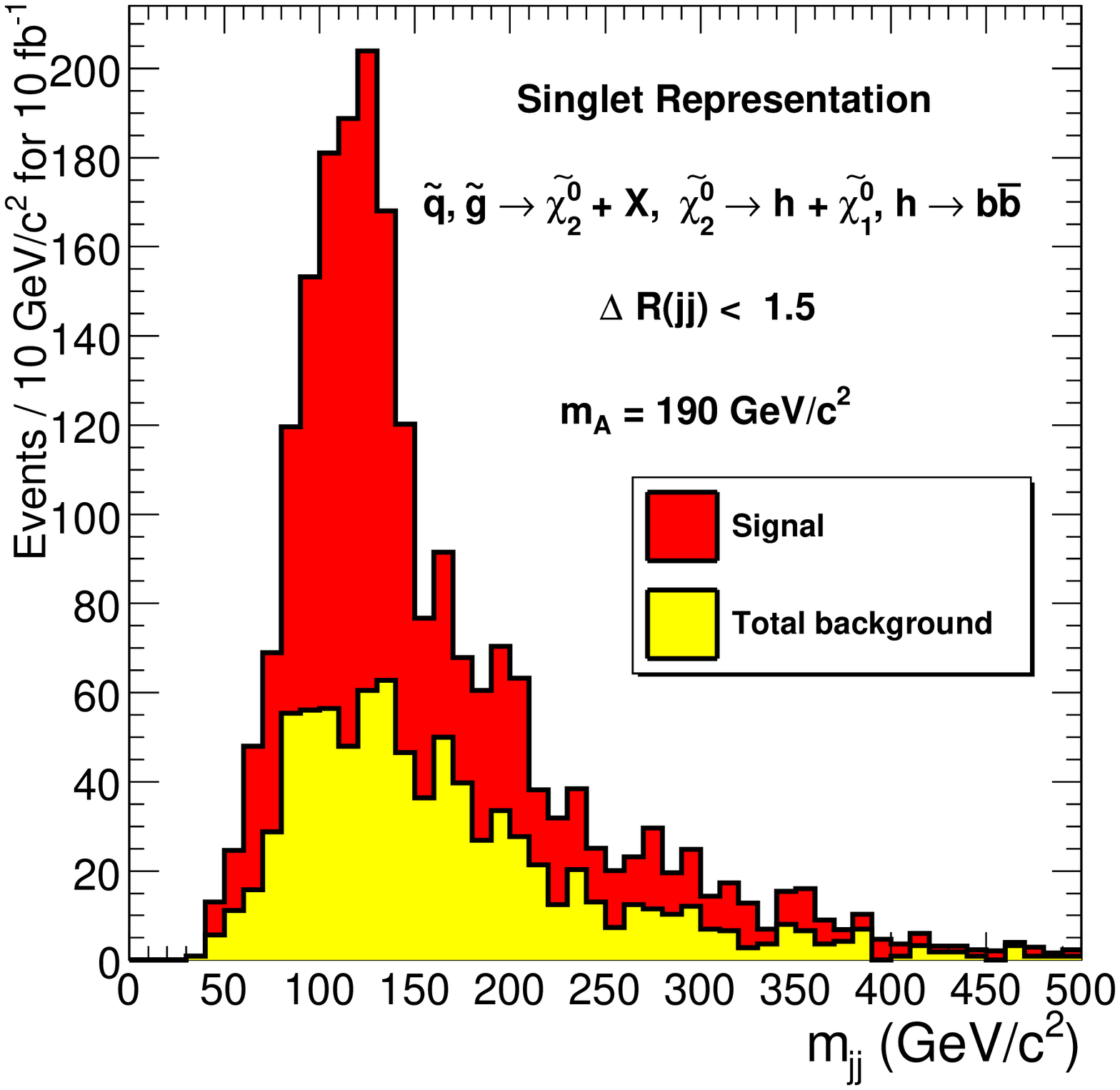}} 
  \caption{Distribution of the invariant mass reconstructed from the selected b jets
for the signal, for the total background with $\Delta \rm R(jj) <$~1.5
in the singlet representation with m$_{\rm A}$~=~190~GeV/$c^2$. }
  \label{fig:hmassRep1DR}
  \end{minipage}
\\
  \begin{minipage}{7.5cm}
    \centering
  \resizebox{\linewidth}{60 mm}{\includegraphics{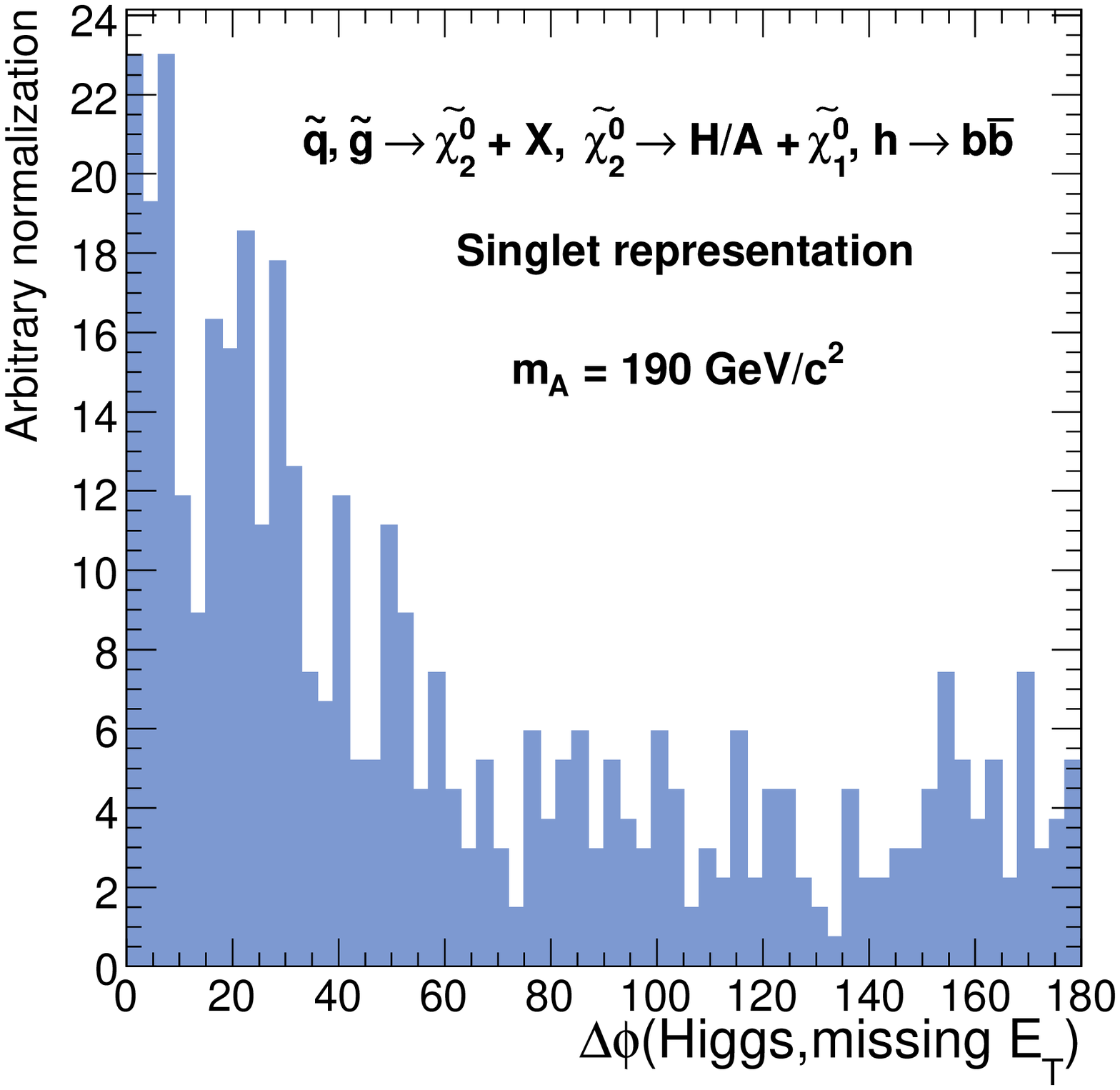}}
  \caption{Distribution of $\Delta\phi$ between the Higgs boson reconstructed from the selected b jets
and missing $\rm E_{\rm T}$ for reconstructed missing $\rm E_{\rm T}$ (solid histogram)
and for missing $\rm E_{\rm T}$ reconstructed at the generator level (solid line)
in the singlet representation with m$_{\rm A}$~=~190~GeV/$c^2$.}
  \label{fig:deltaPhiRep1}
  \end{minipage}
  &
  \begin{minipage}{7.5cm}
    \centering
  \resizebox{\linewidth}{60 mm}{\includegraphics{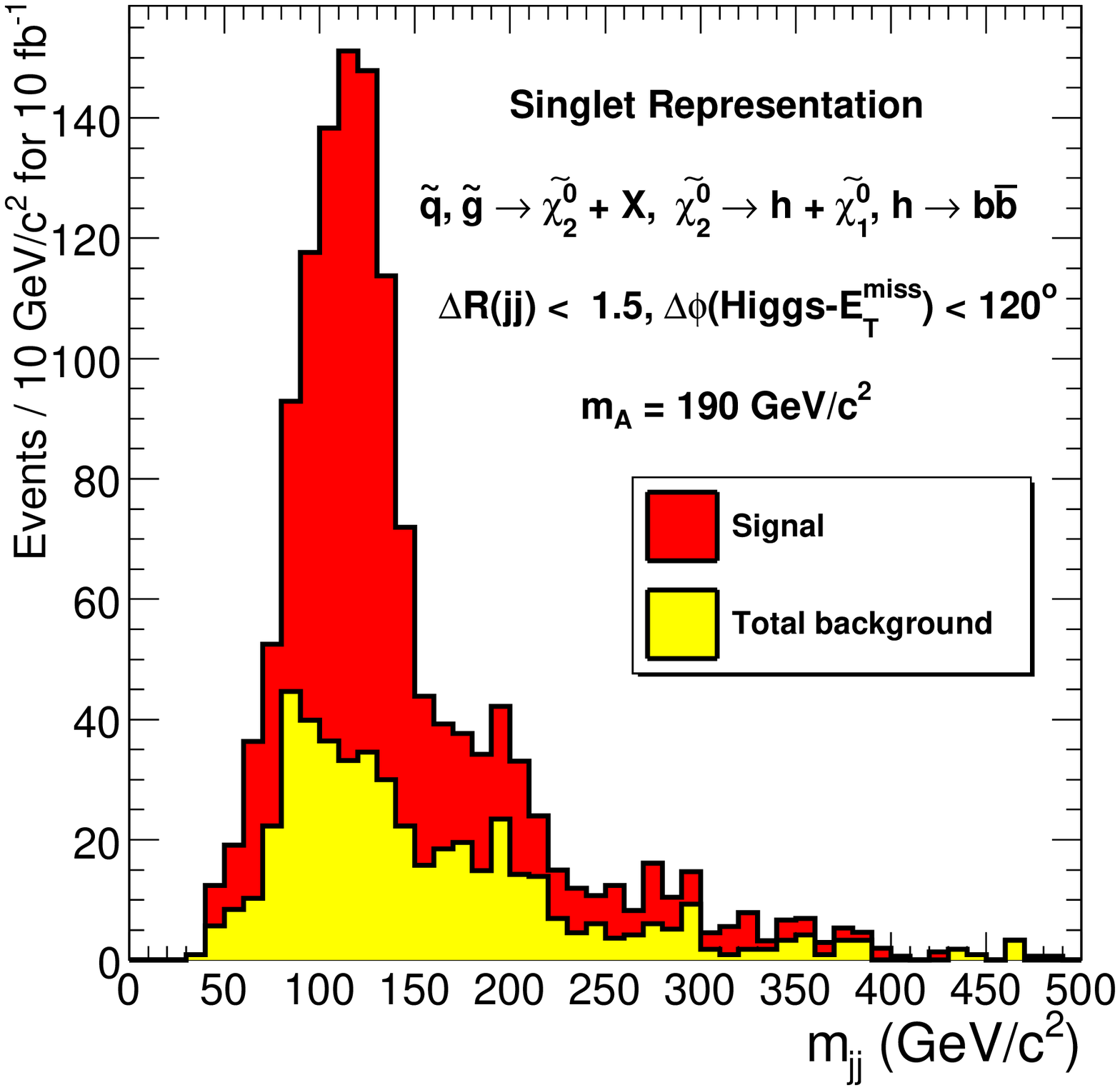}} 
  \caption{Distribution of the invariant mass reconstructed from the selected b jets
for the signal, for the total background with $\Delta \rm R(jj) <$~1.5
and $\Delta\phi(\rm Higgs,\rm E_{\rm T}^{\rm miss} <$~120$^0$)
in the singlet representation with m$_{\rm A}$~=~190~GeV/$c^2$.\newline }
  \label{fig:hmassRep1DRphi}
  \end{minipage}
  \end{tabular}
\end{figure*}

Figure \ref{fig:hmassRep1} shows the $\rm b\overline{\rm b}$ mass distribution for the signal in the singlet
representation with m$_{\rm A}$~=~190~GeV/$c^2$ and tan$\beta$~=~10 and for the total background. The 
$\rm t\overline{\rm t}$ background is shown separately in the figure. 
Setting an upper bound of $\Delta \rm R <$~1.5 could reduce this tail as can be seen from Fig.~\ref{fig:hmassRep1DR}.
To further reduce the backgrounds correlations between the reconstructed $\rm b\overline{\rm b}$ pair, 
presenting the Higgs boson,
and the missing $\rm E_{\rm T}$ was studied. The missing $\rm E_{\rm T}$ originates from the  
$\tilde{\chi}_{2}$ decays at the end of the two cascades. Therefore,
the missing $\rm E_{\rm T}$ can be expected to follow one of the hemisphere axis. It was found to follow 
preferentially the selected $\rm b\overline{\rm b}$ pair, as can be seen from Fig.~\ref{fig:deltaPhiRep1} showing the 
$\Delta\phi$ distribution between the  $\rm b\overline{\rm b}$ system and the missing $\rm E_{\rm T}$. 
Figure \ref{fig:hmassRep1DRphi} shows the invariant mass distribution with $\Delta\phi(\rm Higgs,\rm E_{\rm T}^{\rm miss}) <$~120$^0$. 

\begin{figure*}
  \centering
  \vskip 0.1 in
  \begin{tabular}{ccc}
  \begin{minipage}{7.5cm}
    \centering
  \resizebox{\linewidth}{60 mm}{\includegraphics{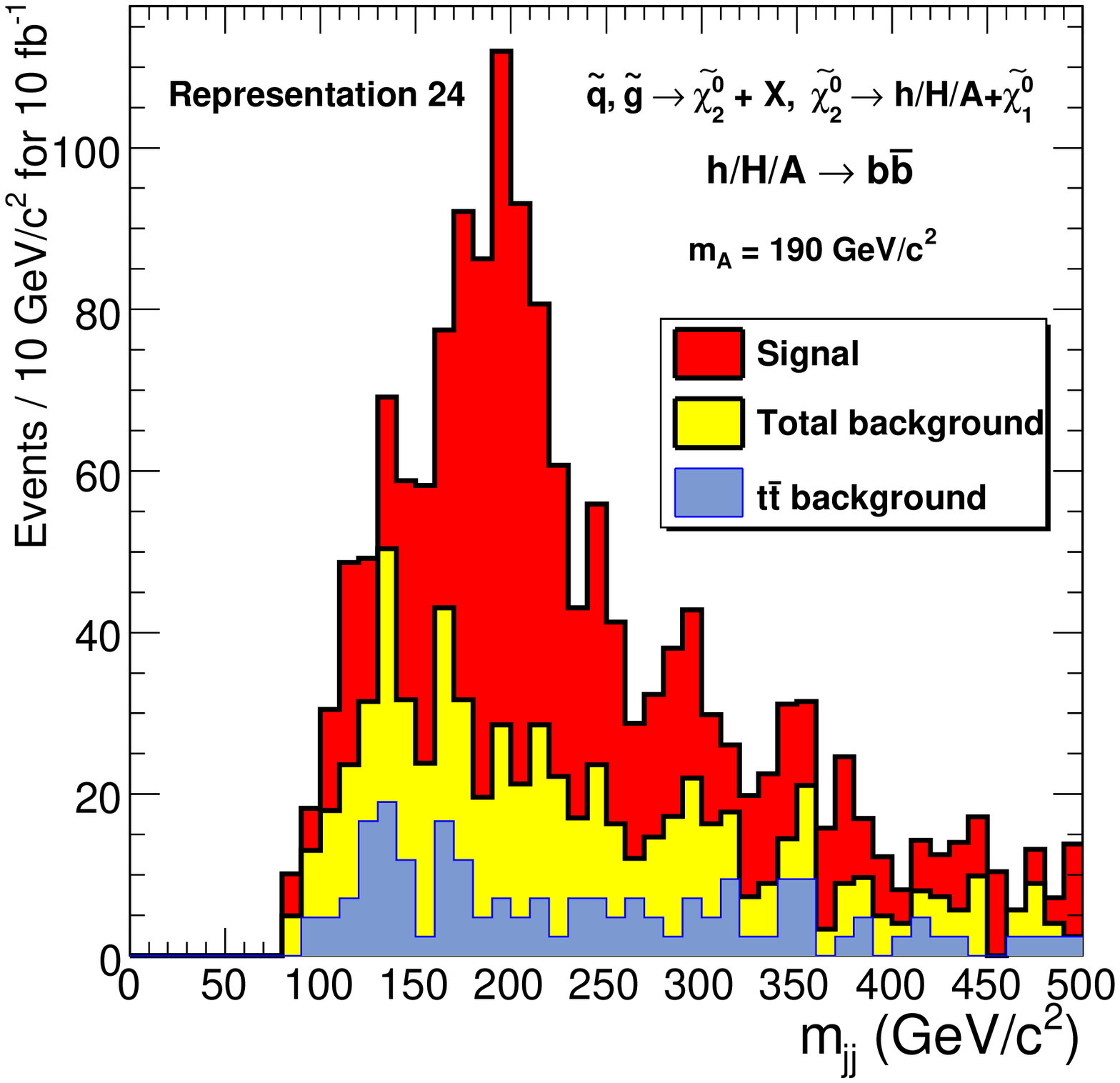}}
  \caption{Distribution of the invariant mass reconstructed from the selected b jets
for the signal, for the total background and for the $\rm t\overline{\rm t}$
background in the representation {\bf 24} for m$_{\rm A}$~=~190~GeV/$c^2$
with  $\rm E_{\rm T} > 100$~GeV for the tagged b jets. }
  \label{fig:HAmass190b100}
  \end{minipage}
  &
  \begin{minipage}{7.5cm}
    \centering
  \resizebox{\linewidth}{60 mm}{\includegraphics{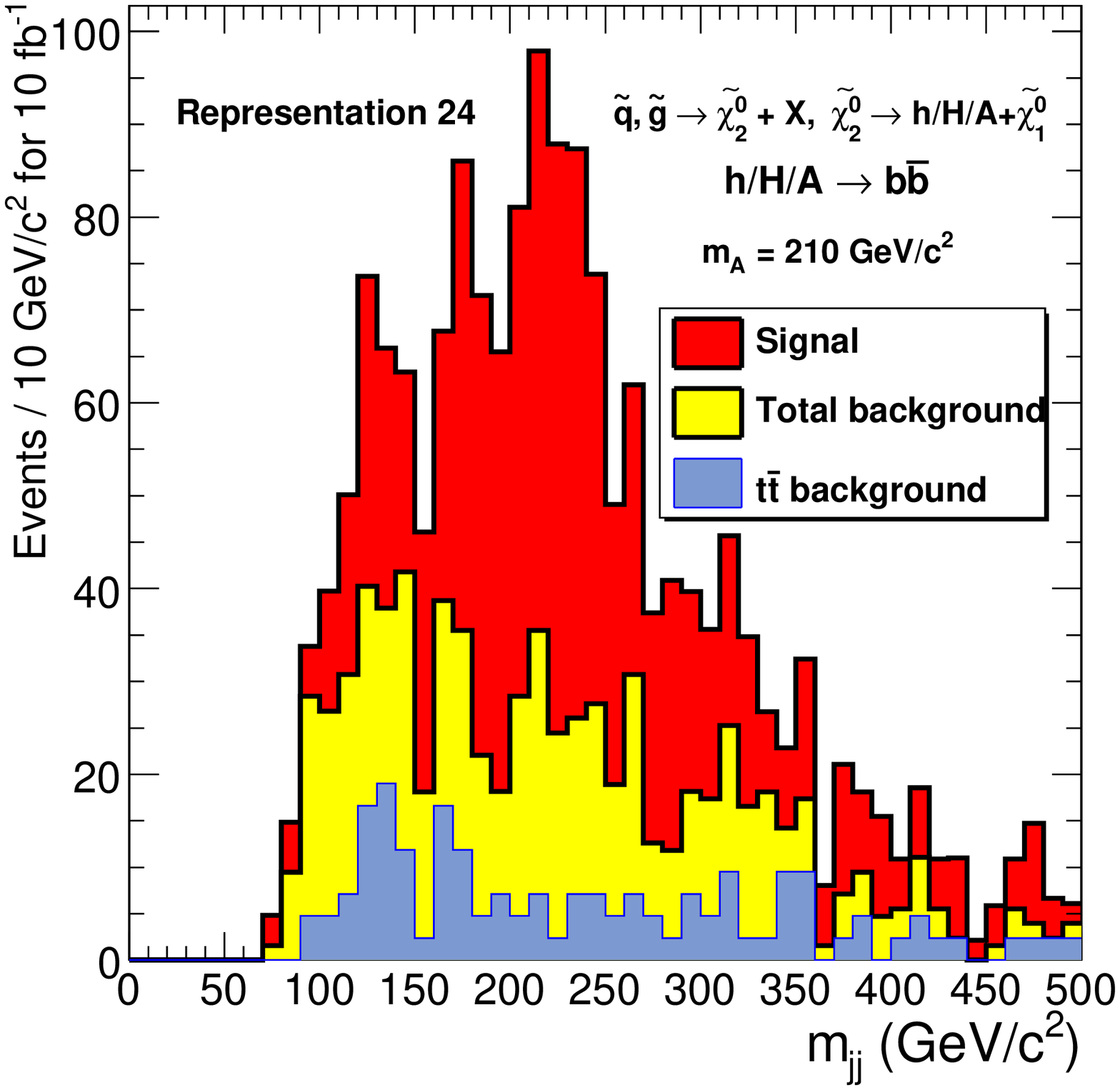}}
  \caption{Distribution of the invariant mass reconstructed from the selected b jets
for the signal, for the total background and for the $\rm t\overline{\rm t}$
background in the representation {\bf 24} for m$_{\rm A}$~=~210~GeV/$c^2$
with  $\rm E_{\rm T} > 100$~GeV for the tagged b jets.}
  \label{fig:HAmass210b100}
  \end{minipage}
  \end{tabular}
\end{figure*}

Figures \ref{fig:HAmass190b100} and \ref{fig:HAmass210b100} show the invariant mass distributions
 for the signal and background in the {\bf 24}-dimensional representation
  with m$_{\rm A}$~=~190 and 210~GeV/$c^2$ and with $\rm E_{\rm T}^{b jet}>$~100~GeV. 

\begin{table*}
\small
\centering
\vskip 0.1 in
\caption{Cross section times branching fraction, number of events for 10 fb$^{-1}$ and percentage cut efficiencies (in brackets) for 
the H/A signal with m$_{\rm A}$~=~190~GeV/$c^2$ and for the SUSY background in the {\bf 24} dimensional representation
with tan$\beta$~=~10.
 The results for the $\rm t\bar{\rm t}$ and QCD backgrounds are also shown. For the QCD background the efficiency 
for $\rm E_{\rm T}^{\rm Miss} >$ 200 GeV was
calculated with full simulation \cite{cmssw}, without trigger simulation.}
\begin{tabular}{|l|cc|cc|cc|cc|}
  \hline
\large rep 24				& \multicolumn{2}{|c|}{Signal}	
                                        & \multicolumn{2}{|c|}{SUSY}	
                                        & \multicolumn{2}{|c|}{$\rm t\bar{\rm t}$}
                                        & \multicolumn{2}{|c|}{QCD}      \\
  \hline
  \hline
   $\sigma \times \rm BR$ (pb)     		& 5.07	&       & 18.05 &  	& 840 & 	&$6.29*10^{4}$&\\ 
  \hline
   L1+HLT                          		& 36118	&(71.2) & $1.51*10^{5}$ &(83.7) 	& $3.16*10^{5}$ & (3.8)	& -  &   \\
  \hline
   At least 4 jets, $\rm E_{\rm T} > 50$ GeV  	& 23232	&(64.3) & $1.28*10^{5}$ &(84.6) & $2.54*10^{5}$ & (80.4)	& $1.96*10^{9}$&   \\ 
  \hline
   $\rm E_{\rm T}^{\rm Miss} >$ 200 GeV  	& 22871	&(98.5) & 96382 &	(75.4)  & $48831$ & (19.2)	& 233949 & (0.012)\\ 
   \hline
   $\rm E_{\rm T}^{\rm jet}$ cuts               & 17315	&(75.7) & 53558 &	(55.6)  & 17352 &(35.5)		&15334& (6.6) \\ 
   \hline
   At least 2 b jets, $\rm E_{\rm T} > 100$ GeV & 4663 &(26.9)	& 4335 & (8.1 )	& 3528 &(20.3)		& 184 & (1.2) \\ 
   \hline
   Jet pairing selection               		& 948 &(20.3)   & 503 & (11.6)	   	& 292 &(8.3)		& 4.4 & (2.4) \\ 
   \hline
  
\end{tabular}
\label{table:Rep24}
\end{table*}

\begin{table*}
\small
\centering
\vskip 0.1 in
\caption{Cross section times branching fraction, number of events for 10 fb$^{-1}$ and prosentual cut efficiencies (in brackets) for 
the h signal with m$_{\rm A}$~=~190~GeV/$c^2$ and for the SUSY background in the singlet representation with tan$\beta$~=~10.
The results for the $\rm t\bar{\rm t}$ are also shown.}
\begin{tabular}{|l|cc|cc|cc|cc|}
  \hline
\large rep 1				& \multicolumn{2}{|c|}{Signal}	
                                        & \multicolumn{2}{|c|}{SUSY}	
                                        & \multicolumn{2}{|c|}{$\rm t\bar{\rm t}$} \\
  \hline
  \hline
  \hline
   $\sigma \times \rm BR$ (pb)     		& 4.13	&        & 14.2   &  	& 840 & 	\\ 
  \hline
   L1+HLT                          		& 31782 &(64.8)  & 119582    &(50.4) 	&  $3.16*10^{5}$ & (3.8)	\\
  \hline
   At least 4 jets, $\rm E_{\rm T} > 50$ GeV    & 30682 &(96.9)  & 96548 &(80.7)   	& $2.54*10^{5}$ & (80.4) \\ 
  \hline
   $E_{t}^{Miss} >$ 200 GeV              	& 20627 &(67.2)  & 73157 & (75.8)  	& $48831$ & (19.2)	\\ 
   \hline
   $E_{t}^{jet}$ cuts                  		& 15315 &(74.2)  & 48260 & (66.0)  	& 17352 &(35.5)	   	\\ 
   \hline
   At least 2 b jets, $\rm E_{\rm T} > 50$ GeV  & 5494 &(35.9)	 & 5570 & (11.5)	& 5098 &(29.4)		\\ 
   \hline
   Jet pairing selection               		& 1813 &(33.0)   & 839 & (15.1)		& 606 &(11.9)		\\ 
   \hline
  $\Delta\phi(\rm Higgs,\rm E_{\rm T}^{\rm miss}) <$~120$^o$	& 866 & (69.5)		 & 254 & (51.6)		& 235 & (61.8)			\\
   \hline
 
\end{tabular}
\label{table:Rep1}
\end{table*}

The selection cuts are summarized below:
\begin{itemize}

        \item At least 4 reconstructed jets with $\rm E_{\rm T} > 50$ GeV

        \item One jet with $\rm E_{\rm T} > 200$ GeV

        \item Two jets with $\rm E_{\rm T} >150$ GeV

        \item Three jets with $\rm E_{\rm T} > 100$ GeV

        \item At least 2 b jets with $\rm E_{\rm T} > 50$ GeV for the light Higgs boson h and
             at least 2 b jets with $\rm E_{\rm T} > 100$ GeV for the heavy Higgs bosons

        \item Missing $\rm E_{T}$ over 200 GeV

\end{itemize}

 The number of events after each cut and relative
cut efficiencies are shown in Tables \ref{table:Rep24} and
\ref{table:Rep1}. The QCD background was estimated without trigger simulation and 
with full simulation for the efficiency of the $\rm E_{\rm T}^{\rm miss}>$~200~GeV cut.

\vspace{1ex}

\section{Estimation of statistical significance}

\begin{figure*}
  \centering
  \vskip 0.1 in
  \begin{tabular}{ccc}
  \begin{minipage}{7.5cm}
    \centering
  \resizebox{\linewidth}{60 mm}{\includegraphics{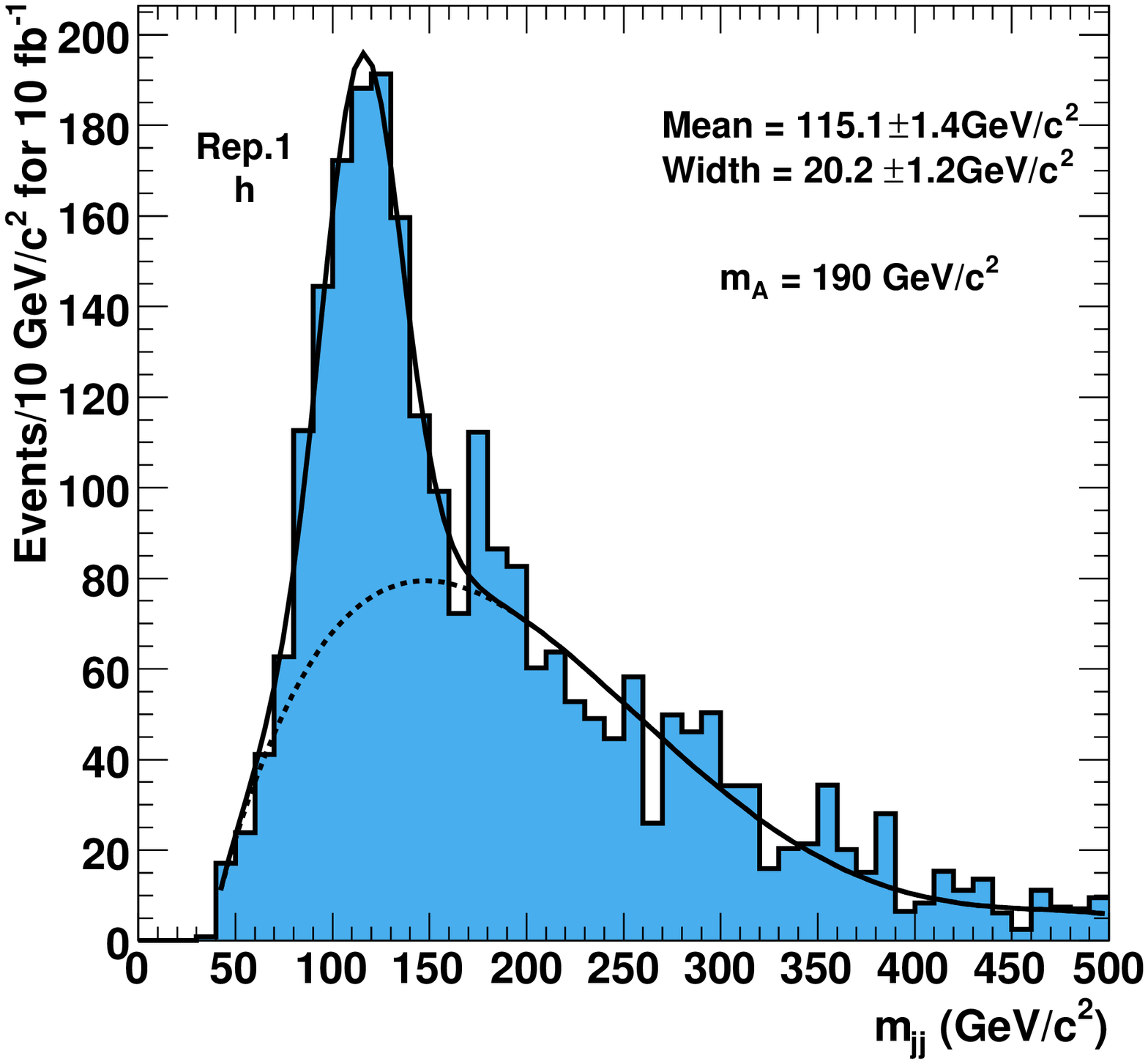}}
  \caption{Fit of a Gaussian plus a 5th degree polynomial to the $\rm b\overline{\rm b}$ invariant mass
for the signal and the total background in the singlet representation with m$_{\rm A}$~=~190~GeV/$c^2$. }
  \label{fig:fitRep1}
  \end{minipage}
  &
  \begin{minipage}{7.5cm}
    \centering
  \resizebox{\linewidth}{60 mm}{\includegraphics{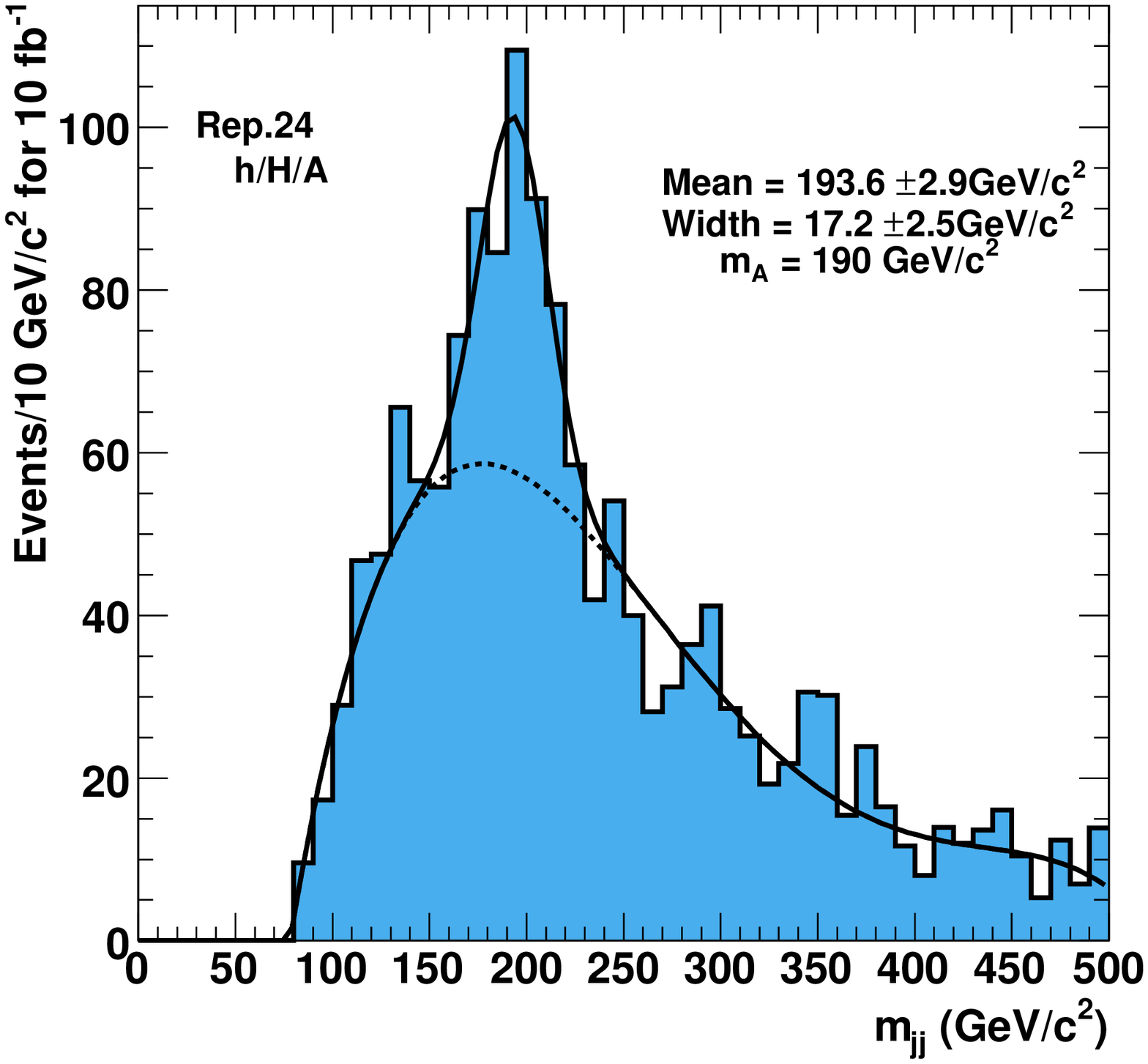}}
  \caption{Fit of a Gaussian plus a 5th degree polynomial to the $\rm b\overline{\rm b}$ invariant mass
for the signal and the total background in the {\bf 24}-representation with m$_{\rm A}$~=~190~GeV/$c^2$.}
  \label{fig:fit_mA190rep24}
  \end{minipage}
  \end{tabular}
\end{figure*}
\begin{figure*}
  \centering
  \vskip 0.1 in
  \begin{tabular}{ccc}
  \begin{minipage}{7.5cm}
    \centering
  \resizebox{\linewidth}{60 mm}{\includegraphics{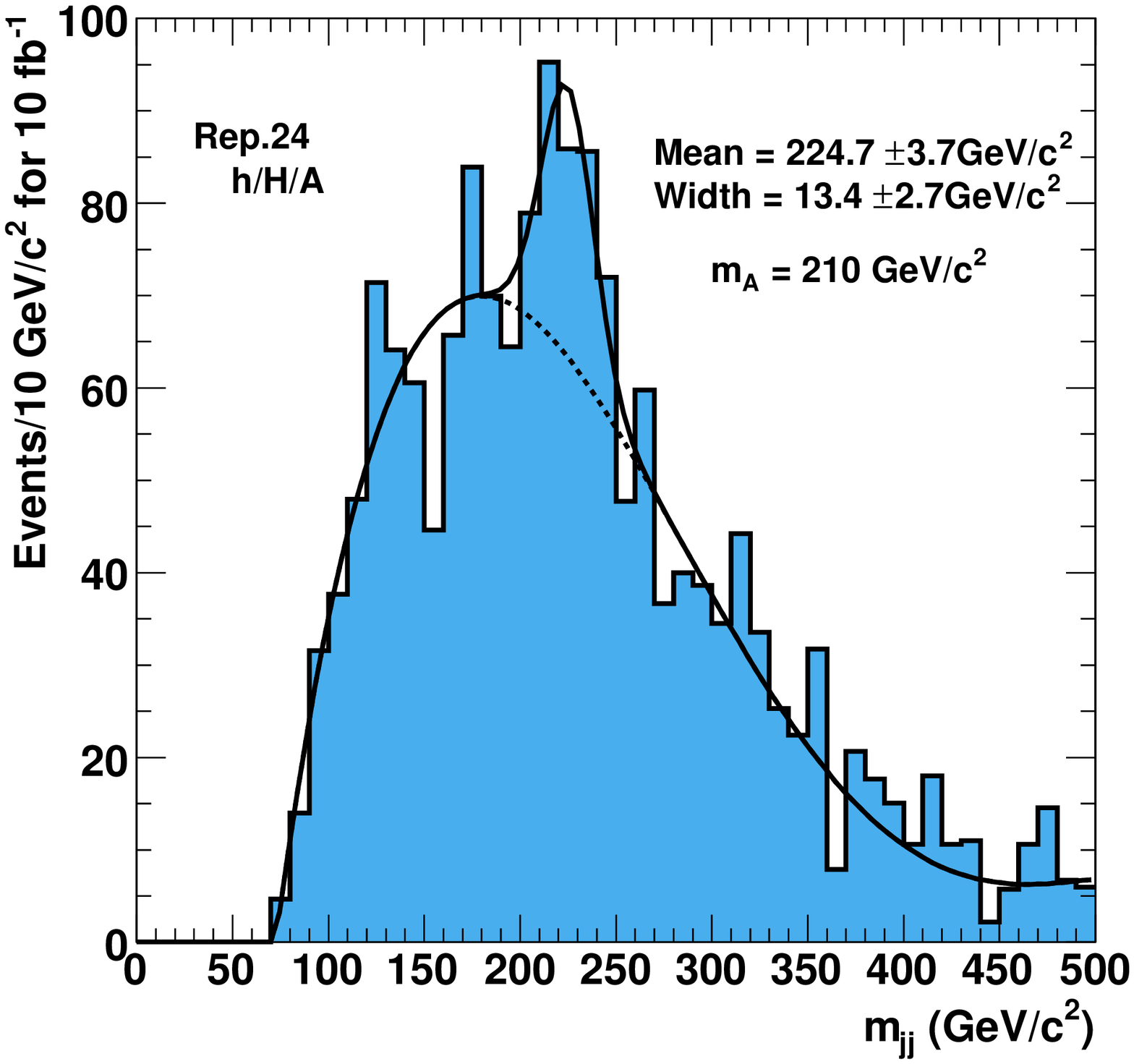}}
  \caption{Fit of a Gaussian plus a 5th degree polynomial to the $\rm b\overline{\rm b}$ invariant mass
for the signal and the total background in the {\bf 24}-representation with m$_{\rm A}$~=~210~GeV/$c^2$. }
  \label{fig:fit_mA210rep24}
  \end{minipage}
  &
  \begin{minipage}{7.5cm}
    \centering
  \resizebox{\linewidth}{60 mm}{\includegraphics{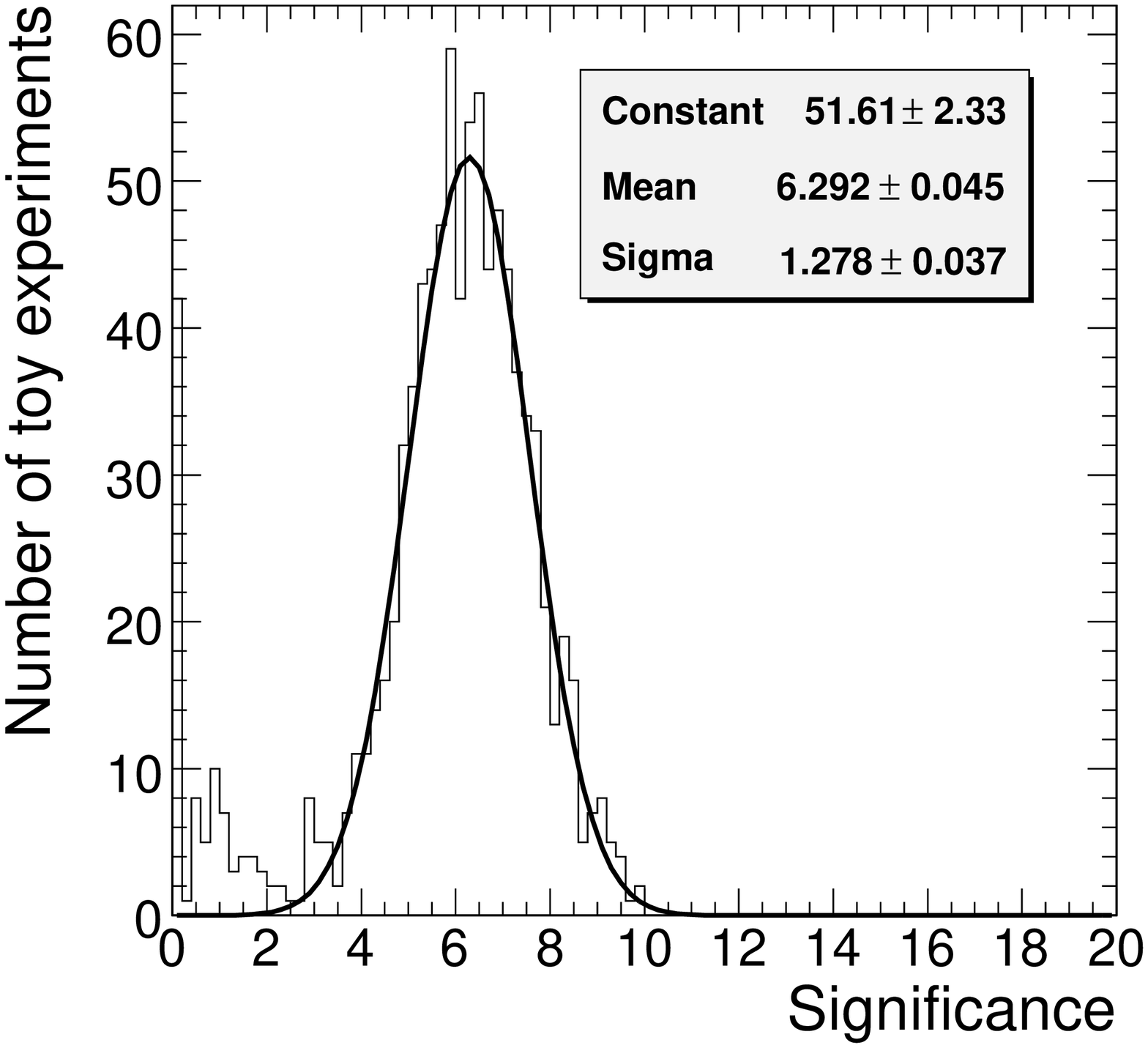}}
  \caption{Distribution of the likelihood-ratio estimator $\sigma_{\it L}$
for $\tilde{\rm q},\tilde{\rm g} \ra \tilde{\chi}_{1}\rm H/A \ra
\tilde{\chi}_{1}\rm b\overline{\rm b}$ in the representation {\bf 24} for m$_{\rm A}$~=~210~GeV/$c^2$.\newline}
  \label{fig:signifDistribution}
  \end{minipage}
  \end{tabular}
\end{figure*}

The event selection procedure described above yields a visible signal in the
 $\rm b\bar{\rm b}$ mass as can be seen from Figs.~\ref{fig:hmassRep1} - \ref{fig:HAmass210b100}.
Therefore the background shape can be determined with fits
to the side-bands of the peak. Approximately 50\% of the background
level in the signal area is due to the combinatorial background.
For the light and heavy Higgs bosons with m$_{\rm A}$~=~190~GeV/$c^2$
in the {\bf 24}- and {\bf 1}-dimensional representations the background areas were taken as those
outside the intervals of 120~$<\rm m_{\rm b\bar{\rm b}} <$~250~GeV/$c^2$ and
80~$<\rm m_{\rm b\bar{\rm b}} <$~150~GeV/$c^2$, respectively. An interval of
150~$<\rm m_{\rm b\bar{\rm b}} <$~250~GeV/$c^2$
was used for m$_{\rm A}$~=~210~GeV/$c^2$ in the {\bf 24}-dimensional representation.
The background distributions were obtained fitting a 5$^{th}$-order polynomial to the
sidebands of the mass peaks, the areas outside the mass windows given above.
The full mass distributions were then fitted combining the obtained polynomial and a Gaussian
distribution. The expected number of background events was extracted from extrapolations of the
background fits to the signal area. The signal events are then given as the difference between
the measured number of events and the expected number of background events from the fit.
Table \ref{table:massfits} shows the mass intervals, the fitted Higgs boson mass values and the resulting
statistical significance. 
Figures \ref{fig:fitRep1} - \ref{fig:fit_mA210rep24} show the fitted mass distributions for the signal
and backgrounds. 

\begin{table*}
\centering
\caption{Higgs boson mass window used for the fits, Higgs boson mass and statistical significance 
obtained from fits with likelihood-ratio method starting from the shapes of the fitted distributions.}
\begin{tabular}{|l|c|c|c|}
\hline
	& rep24, m$_{\rm A}$~=~190~GeV/$c^2$ & rep24, m$_{\rm A}$~=~210~GeV/$c^2$ &   rep1,m$_{\rm A}$~=~190~GeV/$c^2$ 	\\
\hline
\hline
$\Delta$m (GeV/$c^2$)	& 150 - 250	& 150 - 250	& 70 - 150	\\
$\rm m_{\rm Higgs}$ (GeV/$c^2$)	& 193.6$\pm$2.9	& 224.7$\pm$3.7	&	118.5$\pm$1.5	\\
Significance  			& 9.5		& 	5.6	&	21.2	\\
\hline
\end{tabular}
\label{table:massfits}
\end{table*}

To estimate the statistical significance a likelihood-ratio method was used starting from the 
shapes of the fitted distributions for the signal+background and for background alone. 
A new ``toy experiment'' was formed fluctuating each bin in the distribution according to a Poisson 
distribution with the smeared bin value as a mean. The new distribution was refitted with the 
signal (Gaussian) and background (5th order polynomial) shapes. This procedure was repeated 
many (1000) times. For each experiment, a statistical significance was calculated according to
 the following likelihood-ratio estimator:
\begin{eqnarray}
\sigma_{\it L} = \sqrt{2ln(\it L_{\rm S+B} / \it L_{\rm B})} 
\end{eqnarray}

where $\it L_{\rm S+B}$ is the maximum likelihood value obtained in the signal+background binned 
likelihood fit, and  $\it L_{\rm B}$ is the maximum likelihood value from the binned background-only hypothesis fit.
Figure \ref{fig:signifDistribution} shows the distribution of $\sigma_{\it L}$ 
for m$_{\rm A}$~=~190~GeV/$c^2$ in the {\bf 24}-dimensional representation with 10~fb$^{-1}$. The distribution
has a Gaussian shape with width close to unity. The statistical significance can be therefore obtained
as a mean value of a Gaussian fit and is 8.6 for m$_{\rm A}$~=~190~GeV/$c^2$.
The same procedure was performed for  m$_{\rm A}$~=~210~GeV/$c^2$.
The fitting procedure was found to be quite robust and not sensitive to the chosen signal region.
No systematic uncertainties are included. The background shape and the form of the function to fit it
was assumed to be determined from MC studies. 
The SM backgrounds will be ultimately measured from the data.

\section{Methods to measure the backgrounds from data}

Methods to measure the SM backgrounds from data have been developed in CMS \cite{pTDR2}. 
Missing  $\rm E_{\rm T}$ in the QCD multi-jet events is mainly due to detector resolution and mis-measured jets.
Contribution from badly measured jets can be suppressed with topological cuts, excluding events where the 
missing  $\rm E_{\rm T}$ is parallel to the leading jet. For the QCD background, cleaned as much as possible from the
mis-measured jets, the shape and normalization could be obtained measuring the multi-jet events with low $\rm E_{\rm T}$
thresholds, where the SUSY contribution is insignificant, and extrapolating the $\rm E_{\rm T}^{\rm Miss}$ distribution 
to the area of the SUSY searches. Due to large cross section of the QCD multi-jet background the contamination from 
 $\rm t\overline{\rm t}$ events is small and could be suppressed demanding that no b-jets are present in the sample.

The large $\rm E_{\rm T}^{\rm Miss}$ values in the $\rm t\overline{\rm t}$  events with hadronic selection are
due to detector resolution, mis-measured jets, un-identified leptons from W decays, semileptonic 
decays of b and c quarks and hadronic $\tau$ decays from $\rm W \ra \tau\nu$. As for the QCD multi-jet background
the contribution from mis-measured jets can be suppressed with topological cuts. The hadronic $\tau$ decays and
the leptons in jets can be efficiently identified with the newly developed Particle Flow methods \cite{PFlow}. Requirement
of two well identified b jets and a well measured hadronic top mass may not be, however, enough to suppress a contamination
from QCD multi-jet background for the selection of $\rm t\overline{\rm t}$  events with low  $\rm E_{\rm T}$
thresholds. Therefore, the hadronic $\rm t\overline{\rm t}$ background may have to be estimated with Monte-Carlo methods.

\section{Discovery potential}

\begin{figure}[h]
  \centering
  \includegraphics[width=0.46\textwidth]{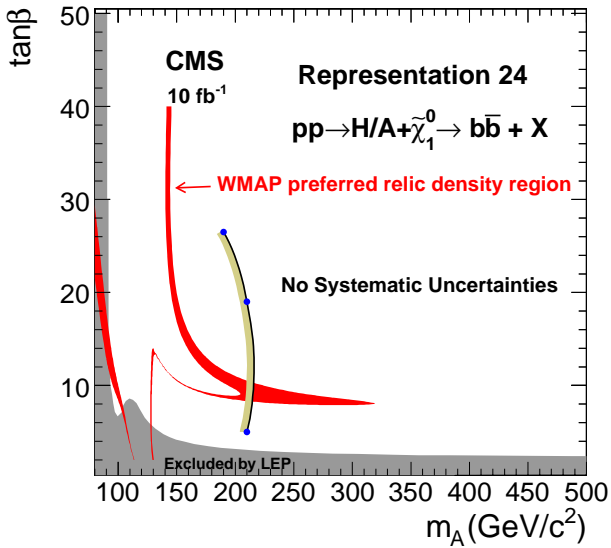}
  \caption{CMS discovery potential at 5$\sigma$ level for $\tilde{\rm q},\tilde{\rm g}
\ra \tilde{\chi}_{1}\rm H/A \ra \tilde{\chi}_{1}\rm b\overline{\rm b}$
in the representation {\bf 24} with an integrated luminosity of 10~fb$^{-1}$.
No systematic uncertainties are included.}
  \protect\label{fig:discovery_rep24}
\end{figure}

Figure \ref{fig:discovery_rep24} shows the 5$\sigma$-discovery potential 
for Higgs boson production in the SUSY cascades 
$\tilde{g} \ra \tilde{q}q$, $\tilde{q} \ra  \tilde{\chi_2}q$,
$ \tilde{\chi_2} \ra \tilde{\chi_1} h/H/A$, $h/H/A  \ra b\bar{b}$ 
 in the non-universal {\bf 24} 
dimensional representation for integrated luminosity of 10~fb$^{-1}$. 
Two mass points, $\rm m_{\rm A}$~=~ 190 and 210~GeV/$c^2$ were investigated. The discovery
potential was scanned by renormalizing the signal and SUSY background cross sections as a function
of tan$\beta$. The cross sections were calculated as described in Section 3.
Constant signal and background selection efficiency as a function of tan$\beta$ was assumed.
 
The discovery region extends to masses of $\sim$~210~GeV/$c^2$ around tan$\beta$~=~10 and
to $\sim$~190~GeV/$c^2$ around tan$\beta$~=~30. For $\sim$~190~GeV/$c^2$ the low tan$\beta$
region was covered with better than 5$\sigma$ significance down to tan$\beta$ = 2 which was
the lowest value of tan$\beta$ studied.
Since the 
cross section for H/A production decreases rapidly at Higgs masses around 
$\sim$~280~GeV/$c^2$, the discovery region is not expected to cover the 
high Higgs mass values of (m$_{\rm A}$,tan$\beta$) parameter space
even with high luminosities.
However, large fraction of the WMAP preferred relic
density region can be explored with this channel at the LHC already with 10~fb$^{-1}$.

The systematic uncertanties are not included in the result of Fig.~\ref{fig:discovery_rep24}.
The main source of systematic uncertainties is due the uncertainty uncertainty of the 
jet energy scale, $\sim$~10\% \cite{pTDR2}, which can affect the efficiency of selecting the 
4 jets and the efficiency of the $\rm E_{\rm T}^{\rm Miss}$ cut. Further sources of systematic
 uncertainty are the b-tagging ($\sim$~5\%) and the luminosity measurements ($\sim$~5\%) \cite{pTDR2}.
The theoretical uncertainty on the tt cross section due to variation of the scale and PDF is 
about 6\% \cite{pTDR2}. 
The MC statistics of the $\rm t\overline{\rm t}$ background used in this work 
corresponds to 2.4 fb$^{-1}$, which causes an addional uncertainty of about 
4\% on the level of this background after selection.
Including the systematic uncertainties of similar level has been shown in Ref.~\cite{pTDR2} to degrate the discovery
limits in the SUSY parameter scape only slighty for integrated luminosities up to 10~fb$^{-1}$.
In particular, the reach for the jet+MET channel has been shown to vary by $\lsim$~5\% as a function
of m$_0$ and m$_{1/2}$ for 10~fb$^{-1}$. Similar variation ($\sim$~10~GeV/$c^2$) 
can be expected for the reach of Fig.~\ref{fig:discovery_rep24} in $\rm m_{\rm A}$.

\section{Conclusions}

The dark matter allowed regions in the SU(5)
GUT representations were studied. From these representations all but the singlet lead to non-universal
gaugino masses.
It was demonstrated that combinations of different representations can
give observed dark matter for otherwise experimentally allowed
parameter values.
In the representation {\bf 24} the Higgs signal from the neutralino decay is interesting and 
different from the usual universal singlet model. Therefore this work especially concentrated on 
this representation.
The preferred dark matter regions for several 
parameter selections were showed and analyzed, including the points studied with realistic
detector simulations.

To study the discovery potential at the LHC, the Higgs boson, produced in one of the
SUSY cascades, was assumed to decay to b quarks.
 Event selection was based on the 
requirement of large missing $\rm E_{\rm T}$, four energetic jets 
of which two tagged b jets. 
 The combinatorial background from false jet assignments was 
reduced by separating the jets into two hemispheres. Contributions from the SUSY and SM ($t\overline{t}$)
backgrounds were found to be comparable. Visible signals were 
obtained as a function of the $\rm b\overline{\rm b}$ invariant mass. A likelihood-ratio
method was used to extract the statistical significance. The simulation work showed that
a 5$\sigma$-discovery can be expected up to the masses of $\rm m_{\rm A}$~=~190 - 210~GeV/$c^2$ 
for tan$\beta$~=~5-35. The WMAP preferred relic density region, excluding those for  
tan$\beta \sim$~8-10, can be explored at the LHC  in the representation {\bf 24} with the 
$\tilde{g} \ra \tilde{q}q$, $\tilde{q} \ra  \tilde{\chi_2}q$,
$ \tilde{\chi_2} \ra \tilde{\chi_1} h/H/A$, $h/H/A  \ra b\bar{b}$
cascades in SUSY events with the integrated luminosity of 10~fb$^{-1}$. 

\section*{Acknowledgments}

The authors would like to thank U. Berzano, F. Moortgat and A. Nikitenko for helpful 
discussions.
SR acknowledges the kind hospitality of the Helsinki Institute of
Physics during the final stages of this work.
Financial support from the Academy of Finland (Project number 115032)
is acknowledged.
The work of JL is supported in part by Bundesministerium f\"ur
Bildung und Forschung, Berlin-Bonn.


\bibliography{chi2hbb_v10}

\begin{thebibliography}{10}

\bibitem{Cremmer:1982wb}
E.~Cremmer, S.~Ferrara, L.~Girardello, and Antoine Van~Proeyen.
\newblock Coupling supersymmetric yang-mills theories to supergravity.
\newblock {\em Phys. Lett.}, B116:231, 1982.

\bibitem{Ellis:1985jn}
John~R. Ellis, K.~Enqvist, Dimitri~V. Nanopoulos, and K.~Tamvakis.
\newblock Gaugino masses and grand unification.
\newblock {\em Phys. Lett.}, B155:381, 1985.

\bibitem{Drees:1985bx}
Manual Drees.
\newblock Phenomenological consequences of {$N=1$} supergravity theories with
  nonminimal kinetic energy terms for vector superfields.
\newblock {\em Phys. Lett.}, B158:409, 1985.

\bibitem{Anderson:1996bg}
G.~Anderson, C~H. Chen, J~F. Gunion, J~D. Lykken, T.~Moroi, and Y.~Yamada.
\newblock Motivations for and implications of non-universal {GUT}-scale
  boundary conditions for soft {SUSY}-breaking parameters.
\newblock 1996.
\newblock hep-ph/9609457.

\bibitem{Huitu:2005wh}
Katri Huitu, Jari Laamanen, Pran~N. Pandita, and Sourov Roy.
\newblock Phenomenology of non-universal gaugino masses in supersymmetric grand
  unified theories.
\newblock {\em Phys. Rev.}, D72:055013, 2005.

\bibitem{Anderson:2000}
Greg Anderson, Howard Baer, Chih-hao Chen, and Xerxes Tata.
\newblock The reach of fermilab tevatron upgrades for {SU(5)} supergravity
  models with nonuniversal gaugino masses.
\newblock {\em Phys. Rev.}, D61:095005, 2000.

\bibitem{Huitu:2000}
Katri Huitu, Yoshiharu Kawamura, Tatsuo Kobayashi, and Kai Puolamaki.
\newblock Phenomenological constraints on {SUSY SU(5) GUTs} with nonuniversal
  gaugino masses.
\newblock {\em Phys. Rev.}, D61:035001, 2000.

\bibitem{Belanger:2005}
G~Belanger, F~Boudjema, A~Cottrant, A~Pukhov, and A~Semenov.
\newblock {WMAP} constraints on {SUGRA} models with non-universal gaugino
  masses and prospects for direct detection.
\newblock {\em Nucl. Phys.}, B706:411, 2005.

\bibitem{Djouadi:2001}
A~Djouadi, Y~Mambrini, and M~Muhlleitner.
\newblock Chargino and neutralino decays revisited.
\newblock {\em Eur. Phys. J.}, C20:563, 2001.

\bibitem{Chattopadhyay:2005}
Utpal Chattopadhyay, Debajyoti Choudhury, and Debottam Das.
\newblock Large evolution of the bilinear {Higgs} coupling in supersymmetric
  models and reduction of phase sensitivity.
\newblock {\em Phys. Rev.}, D72:095015, 2005.

\bibitem{Chattopadhyay:2001mj}
Utpal Chattopadhyay and Pran Nath.
\newblock b - tau unification, $g(\mu)-2$, the $b \to s + \gamma$ constraint
  and nonuniversalities.
\newblock {\em Phys. Rev.}, D65:075009, 2002.

\bibitem{Bhattacharya:2007dr}
Subhaditya Bhattacharya, AseshKrishna Datta, and Biswarup Mukhopadhyaya.
\newblock {Non-universal gaugino masses: a signal-based analysis for the Large
  Hadron Collider}.
\newblock {\em JHEP}, 10:080, 2007.

\bibitem{Corsetti:2001}
Achille Corsetti and Pran Nath.
\newblock Gaugino mass nonuniversality and dark matter in {SUGRA}, strings and
  {D-brane} models.
\newblock {\em Phys. Rev.}, D64:125010, 2001.

\bibitem{Bertin:2003}
Vincent Bertin, Nezri Emmanuel, and Jean Orloff.
\newblock Neutralino dark matter beyond {CMSSM} universality.
\newblock {\em JHEP}, 02:046, 2003.

\bibitem{Birkedal-Hansen:2003}
Andreas Birkedal-Hansen and Brent~D. Nelson.
\newblock Relic neutralino densities and detection rates with nonuniversal
  gaugino masses.
\newblock {\em Phys. Rev.}, D67:095006, 2003.

\bibitem{Chattopadhyay:2003}
Utpal Chattopadhyay and D~P. Roy.
\newblock Higgsino dark matter in a {SUGRA} model with nonuniversal gaugino
  masses.
\newblock {\em Phys. Rev.}, D68:033010, 2003.

\bibitem{Chattopadhyay:2001va}
Utpal Chattopadhyay, Achille Corsetti, and Pran Nath.
\newblock Supersymmetric dark matter and yukawa unification.
\newblock {\em Phys. Rev.}, D66:035003, 2002.

\bibitem{Cerdeno:2004zj}
D.~G. Cerdeno and C.~Munoz.
\newblock Neutralino dark matter in supergravity theories with non- universal
  scalar and gaugino masses.
\newblock {\em JHEP}, 10:015, 2004.

\bibitem{King:2007vh}
S.~F. King, J.~P. Roberts, and D.~P. Roy.
\newblock Natural dark matter in {SUSY GUTs} with non-universal gaugino masses.
\newblock 2007.

\bibitem{Baer:1992}
H.~Baer, M.~Bisset, X.~Tata, and J.~Woodside.
\newblock {\em Phys. Rev.}, D46:303, 1992.

\bibitem{Datta:2003iz}
Aseshkrishna Datta, Abdelhak Djouadi, Monoranjan Guchait, and Filip Moortgat.
\newblock Detection of {MSSM} {Higgs} bosons from supersymmetric particle
  cascade decays at the {LHC}.
\newblock {\em Nucl. Phys.}, B681:31--64, 2004.

\bibitem{arXiv:0806.2367}
Priyotosh Bandyopadhyay, AseshKrishna Datta, and Biswarup Mukhopadhyaya.
\newblock {Signatures of gaugino mass non-universality in cascade Higgs
  production at the LHC}.
\newblock 2008.
\newblock arXiv:0806.2367.

\bibitem{CMSNote2006/090}
F.~Moortgat, P.~Olbrechts, and L.~Pape.
\newblock Search for a light {Higgs} boson in {SUSY} cascades.
\newblock {\em CMS Note 2006/090}.

\bibitem{Spergel:2003cb}
D.~N. Spergel et~al.
\newblock First year wilkinson microwave anisotropy probe ({WMAP})
  observations: Determination of cosmological parameters.
\newblock {\em Astrophys. J. Suppl.}, 148:175, 2003.

\bibitem{Spergel:2006hy}
D.~N. Spergel et~al.
\newblock {Wilkinson Microwave Anisotropy Probe ({WMAP}) three year results:
  Implications for cosmology}.
\newblock {\em Astrophys. J. Suppl.}, 170:377, 2007.

\bibitem{Allanach:2001kg}
B.~C. Allanach.
\newblock {SOFTSUSY}: A {C++} program for calculating supersymmetric spectra.
\newblock {\em Comput. Phys. Commun.}, 143:305--331, 2002.

\bibitem{Belanger:2001fz}
G.~Belanger, F.~Boudjema, A.~Pukhov, and A.~Semenov.
\newblock {micrOMEGAs}: A program for calculating the relic density in the
  {MSSM}.
\newblock {\em Comput. Phys. Commun.}, 149:103--120, 2002.

\bibitem{Belanger:2004yn}
G.~Belanger, F.~Boudjema, A.~Pukhov, and A.~Semenov.
\newblock {micrOMEGAs}: Version 1.3.
\newblock {\em Comput. Phys. Commun.}, 174:577--604, 2006.

\bibitem{Belanger:2006is}
G.~Belanger, F.~Boudjema, A.~Pukhov, and A.~Semenov.
\newblock {micrOMEGAs2.0}: A program to calculate the relic density of dark
  matter in a generic model.
\newblock {\em Comput. Phys. Commun.}, 176:367--382, 2007.

\bibitem{unknown:2001xy}
Search for charged {Higgs} bosons: Preliminary combined results using {LEP}
  data collected at energies up to 209- {GeV}.
\newblock 2001.
\newblock hep-ex/0107031.

\bibitem{Nath:1994tn}
Pran Nath and R.~Arnowitt.
\newblock b to s gamma decay in supergravity grand unification and dark matter.
\newblock {\em Phys. Lett.}, B336:395--401, 1994.

\bibitem{Nath:1994ci}
Pran Nath and R.~Arnowitt.
\newblock Event rates in dark matter detectors for neutralinos including
  constraints from the b to s gamma decay.
\newblock {\em Phys. Rev. Lett.}, 74:4592--4595, 1995.

\bibitem{Baer:1997jq}
Howard Baer, Michal Brhlik, Diego Castano, and Xerxes Tata.
\newblock $b \to s \gamma$ constraints on the minimal supergravity model with
  large tan(beta).
\newblock {\em Phys. Rev.}, D58:015007, 1998.

\bibitem{Carena:2000uj}
Marcela Carena, David Garcia, Ulrich Nierste, and Carlos E.~M. Wagner.
\newblock $b \to s \gamma$ and supersymmetry with large tan(beta).
\newblock {\em Phys. Lett.}, B499:141--146, 2001.

\bibitem{Degrassi:2000qf}
G.~Degrassi, P.~Gambino, and G.~F. Giudice.
\newblock {B to X/s gamma} in supersymmetry: Large contributions beyond the
  leading order.
\newblock {\em JHEP}, 12:009, 2000.

\bibitem{Barberio:2007cr}
E.~Barberio et~al.
\newblock {Averages of b-hadron properties at the end of 2006}.
\newblock 2007.
\newblock arXiv:0704.3575.

\bibitem{Chan:1997bi}
Kwok~Lung Chan, Utpal Chattopadhyay, and Pran Nath.
\newblock Naturalness, weak scale supersymmetry and the prospect for the
  observation of supersymmetry at the tevatron and at the {LHC}.
\newblock {\em Phys. Rev.}, D58:096004, 1998.

\bibitem{Feng:1999mn}
Jonathan~L. Feng, Konstantin~T. Matchev, and Takeo Moroi.
\newblock Multi-tev scalars are natural in minimal supergravity.
\newblock {\em Phys. Rev. Lett.}, 84:2322--2325, 2000.

\bibitem{Baer:2005ky}
Howard Baer, Tadas Krupovnickas, Stefano Profumo, and Piero Ullio.
\newblock Model independent approach to focus point supersymmetry: From dark
  matter to collider searches.
\newblock {\em JHEP}, 10:020, 2005.

\bibitem{Beenakker:1996ch}
W.~Beenakker, R.~Hopker, M.~Spira, and P.~M. Zerwas.
\newblock Squark and gluino production at hadron colliders.
\newblock {\em Nucl. Phys.}, B492:51--103, 1997.

\bibitem{Muhlleitner:2003vg}
M.~Muhlleitner, A.~Djouadi, and Y.~Mambrini.
\newblock {SDECAY}: A fortran code for the decays of the supersymmetric
  particles in the {MSSM}.
\newblock {\em Comput. Phys. Commun.}, 168:46--70, 2005.

\bibitem{Djouadi:1997yw}
A.~Djouadi, J.~Kalinowski, and M.~Spira.
\newblock {HDECAY}: A program for {Higgs} boson decays in the standard model
  and its supersymmetric extension.
\newblock {\em Comput. Phys. Commun.}, 108:56--74, 1998.

\bibitem{Djouadi:2002ze}
Abdelhak Djouadi, Jean-Loic Kneur, and Gilbert Moultaka.
\newblock {SuSpect}: A fortran code for the supersymmetric and {Higgs} particle
  spectrum in the {MSSM}.
\newblock {\em Comput. Phys. Commun.}, 176:426--455, 2007.

\bibitem{cmsweb}
CMS PRS, \url{http://cmsdoc.cern.ch/cms/PRS/results/susybsm/rep24/rep24.html}.
\newblock {\em Search for {Higgs} Bosons in {SUSY} Cascades in {CMS} and Dark
  Matter with Non-universal Gaugino Masses}, 2008.
\newblock Accessed 12 August, 2008.

\bibitem{pythia}
T~Sjostrand, L~Lonnblad, S~Mrenna, and P~Skands.
\newblock Pythia 6.3 physics and manual.
\newblock {\em LU TP}, 03-38, 2003.
\newblock hep-ph/0308153.

\bibitem{cteq5}
H.~L. Lai et~al.
\newblock {Global {QCD} analysis of parton structure of the nucleon: CTEQ5
  parton distributions}.
\newblock {\em Eur. Phys. J.}, C12:375--392, 2000.

\bibitem{cmsim}
Veikko Karimaki.
\newblock {CMSIM}, {CMS} simulation and reconstruction package.
\newblock Site located at \url{http://cmsdoc.cern.ch/cmsim/cmsim.html}.

\bibitem{cmssw}
{CMS} physics technical design report.
\newblock {\em CERN/LHCC 2006-001}, CMS TDR 8.1, 2006.

\bibitem{alpgen}
M.L. Mangano, M.~Moretti, F.~Piccinini, R.~Pittau, and A.D. Polosa.
\newblock Alpgen, a generator for hard multiparton processes in hard
  collisions.
\newblock {\em JHEP}, 001:35, 2003.

\bibitem{pTDR2}
{CMS} physics technical design report.
\newblock {\em CERN/LHCC 2006-002}, CMS TDR 8.2, 2006.

\bibitem{famos}
{CMS OO} reconstruction.
\newblock FAMOS: CMS Reconstruction Package. Site located at
  \url{http://cmsdoc.cern.ch/famos}.

\bibitem{orca}
{CMS OO} reconstruction.
\newblock ORCA: CMS Reconstruction Package. Site located at
  \url{http://cmsdoc.cern.ch/orca}.

\bibitem{note2006_014}
C~Weiser.
\newblock A combined secondary vertex based b-tagging algorithm in {CMS}.
\newblock {\em CMS Note 2006/014}.

\bibitem{L1_TDR}
The level-1 trigger.
\newblock {\em CERN/LHCC 2000-038}, CMS TDR 6.1, 2000.

\bibitem{DAQ_TDR}
Data acquisition and high-level trigger.
\newblock {\em CERN/LHCC 2002-26}, CMS TDR 6.2, 2002.

\bibitem{hemisphereMethod}
L.~Pape and Filip Moortgat.
\newblock Hemisphere algorithm for separation of {SUSY} cascade chains.
\newblock {\em Physics meeting at CERN 10/04/2006}.

\bibitem{PFlow}
CMS Particle~Flow Group.
\newblock The {CMS} particle flow algorithm.
\newblock {\em CMS Note}.
\newblock (In preparation).

\end{thebibliography}

\end{document}